\documentclass[a4paper,11pt]{article}
\pdfoutput=1 

\usepackage{jcappub}
\usepackage{gensymb}
\usepackage{float}

\title{Detection prospects for heavy WIMP dark matter near supermassive black holes, particularly in M31}

\author{Andrei E. Egorov}
\affiliation{Institute of Physics, University of Belgrade, Pregrevica 118, 11080 Belgrade, Serbia}

\emailAdd{aegorov@runbox.com}

\abstract{This work analyzes the detection prospects for weakly interacting massive particles (WIMPs) in dark matter (DM) density spikes around nearby supermassive black holes (SMBHs) by observations in very high energy gamma-ray band. Such spikes are unique targets, which provide a possibility to discover the basic thermal s-wave annihilating WIMP with any mass up to the theoretical unitarity limit $\approx$ 100 TeV. All relevant SMBHs were checked, and only MW* and M31* were identified as worthwhile objects. Cherenkov Telescope	Array (CTA) sensitivity to heavy WIMPs in M31* was estimated. It was obtained that CTA will be able to probe a major part of TeV-scale WIMP parameter space in case of optimistic spike density configuration in M31*. In certain scenarios, M31* may yield even stronger constraints than MW*. Relevant systematic uncertainties were explored. }

\begin{document}
\maketitle
\flushbottom

\section{\label{sec:i}Introduction and motivation}

DM phenomenon was discovered already almost a century ago \cite{1937ApJ....86..217Z,1936ApJ....83...23S}. However, it still presents a mystery in our Universe -- exact physical nature of DM persists to stay unknown despite many decades of active research. A very wide variety of DM candidates is under consideration -- see e.g. \cite{2024arXiv240601705C} for a very new and comprehensive review. Historically WIMPs represent the most probable and motivated candidate, which originates from supersymmetry \cite{1984NuPhB.238..453E}. One of the main WIMP search strategies is indirect or astrophysical, when we aim to detect WIMP or at least constrain its properties through seeking its annihilation signals. WIMPs are indeed a very wide class of particles. Considering the basic thermally produced s-wave annihilating WIMP, we can outline the following contemporary status of the searches. All employed techniques; including analysis of antimatter in cosmic rays (CRs) \cite{2022ScPP...12..163C}, searches of the prompt gamma rays \cite{2025arXiv250820229L}, as well as the secondary synchrotron emission from WIMP annihilation \cite{2023ecrs.confE.120E}; have probed reliably only relatively light WIMPs with the mass $m_x \lesssim$ 0.1 TeV. This is caused mainly by strong decrease of the intensity of all annihilation signals with WIMP mass increase. However, theoretically WIMPs may have the mass up to approximately 100 TeV \cite{2019PhRvD.100d3029S}! Thus, considering the lower mass bound to be few GeV, we see that all the enormous efforts in indirect DM searches have tested quite small part of the thermal WIMP parameter space so far -- nearly one quarter of the whole mass range in logarithmic measure. But our final goal is to scan the \emph{whole} mass range in order to either discover WIMP at some point or exclude it completely as viable DM candidate.

Further progress is anticipated from upcoming CTA Observatory.\footnote{\url{https://www.ctao.org}} This grand next-generation facility will greatly advance capabilities of very high energy gamma-ray astronomy. Indirect DM searches are one of the key CTA objectives. CTA, also together with Southern Wide-field Gamma-ray Observatory (SWGO)\footnote{\url{https://www.swgo.org}} planned for later, are probably the unique instruments, which provide a possibility to discover heavy WIMPs in foreseeable future. Specifically, the best observational target was deduced to be DM halo of our Galaxy near its center. Thus \cite{2021JCAP...01..057A} conducted the detailed study of CTA sensitivity to generic WIMP parameters for that target. Also recently \cite{2025arXiv250608084A} analyzed the specific higgsino WIMP. The results of these works can be briefly summarized as follows. In a good case scenario of cuspy DM density profile in Milky Way (MW), CTA should be able to probe thermal WIMPs in the mass range $\sim(0.1-10)$ TeV. However, if DM density profile has a substantial core with radius $\sim$ 1 kpc or more, CTA will not be able to detect any WIMPs at all. This renders a fundamental uncertainty, since DM density can not be determined with satisfactory precision in inner few kiloparsecs of galaxies like MW. This is conditioned mainly by the fact of substantial domination of baryonic gravitational potential over that of DM there. SWGO is expected to gather qualitatively same sensitivity in (far) future \cite{2019JCAP...12..061V}. Therefore, we can expect significant progress from CTA and SWGO only in quite optimistic case. Such case is illustrated by \cite[figure 14]{2021JCAP...01..057A}: probes of lighter WIMPs, e.g. Fermi-LAT, smoothly join CTA search of heavier WIMPs covering continuously the values up to $m_x\sim 10$ TeV. But even in this case the heaviest possible WIMPs with $m_x\sim (10-100)$ TeV will remain poorly probed. In the case of less favorable DM density distribution, a major fraction of all TeV-scale WIMPs may be unreachable. Also a wide variety of other systematic uncertainties exists. Such perspective is not very inspiring in the field of DM indirect searches, because eventually huge efforts over many decades may fail to test even basic WIMPs.

And a natural question arises: can we invent any other probes of heavy WIMPs in order to complement limited capabilities of the above methodology and extend further our sensitivity? The main guiding principle here is to look for nearby DM objects with highest possible density, which provide a substantial annihilation rate and, hence, tangible emission intensity even from heaviest WIMPs. However, all such nearby DM halos had been explored already: dwarf MW satellites \cite{2019PhRvD..99l3017H} and Andromeda galaxy M31 \cite{2023JCAP...08..073M}. And all of them appeared to provide even worse sensitivity in comparison with the Galactic center region. Having this, the last resort, which we can consider, is more exotic objects -- DM density spikes in a vicinity of nearby SMBHs. Such spikes are expected to form by local gravitational field of host SMBHs. Very high DM density inside spikes may provide sufficiently bright annihilation signal in prompt gamma-ray photons. This paradigm comprises the subject of my work.

\subsection{\label{sec:rev}Brief review of previous studies of WIMP detection around SMBHs}
The idea to search WIMP annihilation signatures from DM density spikes is quite old and well-developed. The pioneering work about it was probably \cite{1999PhRvL..83.1719G}, where the authors modeled highly-idealistic case of adiabatic SMBH growth and obtained very dense/steep resulting spike. Later this topic was widely developed in various directions: theoretical studies of the spike internal structure \cite{2004PhRvL..92t1304M,2004PhRvL..93f1302G,2007PhRvD..76j3532V,2013PhRvD..88f3522S,2016PhRvD..93l3510S,2022PhRvD.106d3018S,2025PhRvD.112d3025Z}, observational measurements of spike properties \cite{2024MNRAS.527.3196S,2025arXiv250610122S} and indeed derivation of implications for annihilating DM \cite{2014PhRvL.113o1302F,2015PhRvD..92d3510L,2015PhRvL.115w1302S,2020PhRvD.102b3030C,2023PhRvD.108j3042C,2023JCAP...08..063B,2025PhRvD.111k5033P}. And this is not a complete list of papers. We can outline the following brief summary of these extensive studies. On the theoretical side, model predictions for the spike density profile are highly uncertain. Another big uncertainty exists in the knowledge of unperturbed DM density around a galactic center, as was mentioned above. This density value is important too, since it provides a basis for the spike. These and other uncertainties propagate into respective DM constraints, which are derived from observational data on various SMBHs. Finally, this results in very model-dependent implications for DM properties. This was brightly illustrated in e.g. \cite{2023PhRvD.108j3042C}, where M87* was employed for constraining p-wave annihilating DM: the most optimistic spike configuration excluded the thermal WIMPs up to $m_x\sim 10$ TeV, but even modest deviations from the idealized scenario vanish the exclusions for any mass!

Although many studies have been made already, a conclusive picture seems to be missing yet, because majority of the accomplished works focused on particular models and targets. And a specific further strategy seems to be absent currently: particularly, which targets would provide the best sensitivity; and which advancements could be realized with the upcoming new gamma-ray facilities? These points motivated my work.

\subsection{\label{sec:obj}Goals and objectives of this work}
Besides a little and somewhat critical review of the current state of annihilating DM searches near SMBHs, I set the following main objectives for the present work.
\begin{enumerate}
	\item Analyze all nearby SMBHs and isolate the most promising among them. Such analysis does not seem to be conducted before. However, it is highly relevant due to observational limitations in very high energy gamma-ray band, where we aim to look for heavy WIMPs. Namely, imaging atmospheric Cherenkov telescopes (IACTs) have relatively small field of view and require long exposure time. Thus, CTA has the field of view about 6\degree--8\degree~in diameter \cite{CTA}. Therefore, CTA will be doing quite targeted sky observations and will not have the opportunity to survey large sky fraction like e.g. Fermi-LAT. And we need to know in advance one or two SMBHs, which are optimal to target with long exposure in order to obtain the best sensitivity to WIMPs. I.e., CTA may not be able to image deeply many SMBHs.
	\item Among the closest SMBHs, M31* was never considered before (best to the author's knowledge). However, this object looks very promising: it is located at a small distance; has a high mass and, last but not least, is very faint in high-energy emissions \cite{2011ApJ...728L..10L}. The latter implies good sensitivity to potential emission due to DM, since SMBH vicinity is a complicated point-like source, which contains various astrophysical emissions from accretion disk, nearby dense medium with PeVatrons etc. All these competing emission sources inevitably create confusing backgrounds. Thus, M31* should be analyzed very thoroughly.
	\item Estimate CTA sensitivity to WIMP parameters for observations of optimal SMBHs. This is interesting to compare with the sensitivity expected from observations of MW DM halo -- how both will supplement each other.
\end{enumerate}

Here I neglect by relativistic effects and work on the simplest s-wave annihilating WIMP model, which means that the annihilation cross section obeys to Bethe law $\sigma \propto 1/v$, i.e. the product $\sigma v$ is velocity-independent. It is known in particle physics, that not all specific DM candidates have such s-wave leading contribution \cite{2013PhRvD..88a4035K}. However, moving step-by-step in indirect WIMP searches, it would be sensible to test fully the simplest s-wave WIMPs first and then to move to more sophisticated cases. Actually, large SMBHs may provide probably a unique probe for otherwise undetectable p-wave annihilating candidates, since their annihilation cross section is highly suppressed by factor $(v/c)^2$ in usual DM halos, and only SMBHs can accelerate nearby DM particles to relativistic speeds elevating the cross section \cite{2015PhRvL.115w1302S,2023PhRvD.108j3042C}. In more general case, both s- and p-wave contributions are present, and we can parametrize the velocity-averaged annihilation cross section as follows:
\begin{equation}\label{eq:sv}
	\langle \sigma v \rangle = \sigma_0 c_0 + \sigma_1 c_1 \langle v^2 \rangle/c^2,
\end{equation}
where $c_0$ and $c_1$ are some positive constants. In such case it is much harder to derive constraints in comparison with pure s-wave case (when $c_1=0$), because $v$ strongly depends on the distance from particle to SMBH. However, when we derive limits on $\sigma_0 c_0$ in pure s-wave approximation, they are likely conservative; because observational data constrains essentially the left hand side of eq. \eqref{eq:sv}, and subtraction of the contribution with $c_1\neq 0$ from the left hand side would even reduce $\sigma_0 c_0$, strengthening the constraints.

Another simplifying assumption here is absence of effects caused by SMBH rotation. Its influence on the annihilation signal from density spike was studied in \cite{2017PhRvD..96h3014F}: the signal (flux) grows with increase of rotation speed, but mildly -- up to factor $\approx$ 2 at most. This is rather negligible correction in comparison with other model uncertainties. Another thorough work \cite{2015ApJ...806..264S} studied really peculiar effects near SMBH event horizon: particularly, Penrose process, which can extend the gamma-ray spectrum from annihilating DM above the cutoff energy $m_x c^2$. However, it is unclear whether the emission from such a tiny volume around SMBH may bear any observable flux.

The content below is organized as follows: section \ref{sec:calc} describes the general procedure for calculation of the gamma-ray flux from DM density spike, section \ref{sec:all} conducts analysis and comparison of all the relevant nearby SMBHs (task 1 from the list above), section \ref{sec:cta} carries the derivation of main results -- CTA sensitivity to WIMP parameters from observation of M31* (tasks 2 and 3 from the list above), and section \ref{sec:con} sums up the results.

\section{\label{sec:calc}General formalism for WIMP signal calculation}

The prompt gamma-ray flux density from some volume with annihilating DM particles can be calculated by the following well-known equation:
\begin{equation}\label{eq:flux}
	\frac{d\Phi}{dE} = \frac{\langle\sigma v\rangle}{8\pi m_x^2} \frac{dN_\gamma}{dE}(E,m_x)\times \int\limits_{\Omega_\text{PSF}}\int\limits_{\text{LoS}} \rho^2(R) d\Omega dl \equiv \frac{\langle\sigma v\rangle}{8\pi m_x^2} \frac{dN_\gamma}{dE}(E,m_x) \times J,
\end{equation}
where $dN_\gamma/dE$ is the photon spectrum from one annihilation (was taken from conventional database called PPPC4DMID \cite{PPPC,2011JCAP...03..051C,2011JCAP...03..019C}), $\rho(R)$ is DM density at distance $R$ from SMBH (spherically symmetric density distribution is assumed), the integration goes over telescope point spread function (PSF) and along line of sight (LoS). The region of interest on sky does not go beyond PSF; since it was checked, that angular size of \emph{any} DM density spike is smaller than CTA angular resolution at all photon energies, and larger region would result in increase of nuisance background only.

\begin{figure}[t]
\centering
\includegraphics[width=0.45\textwidth]{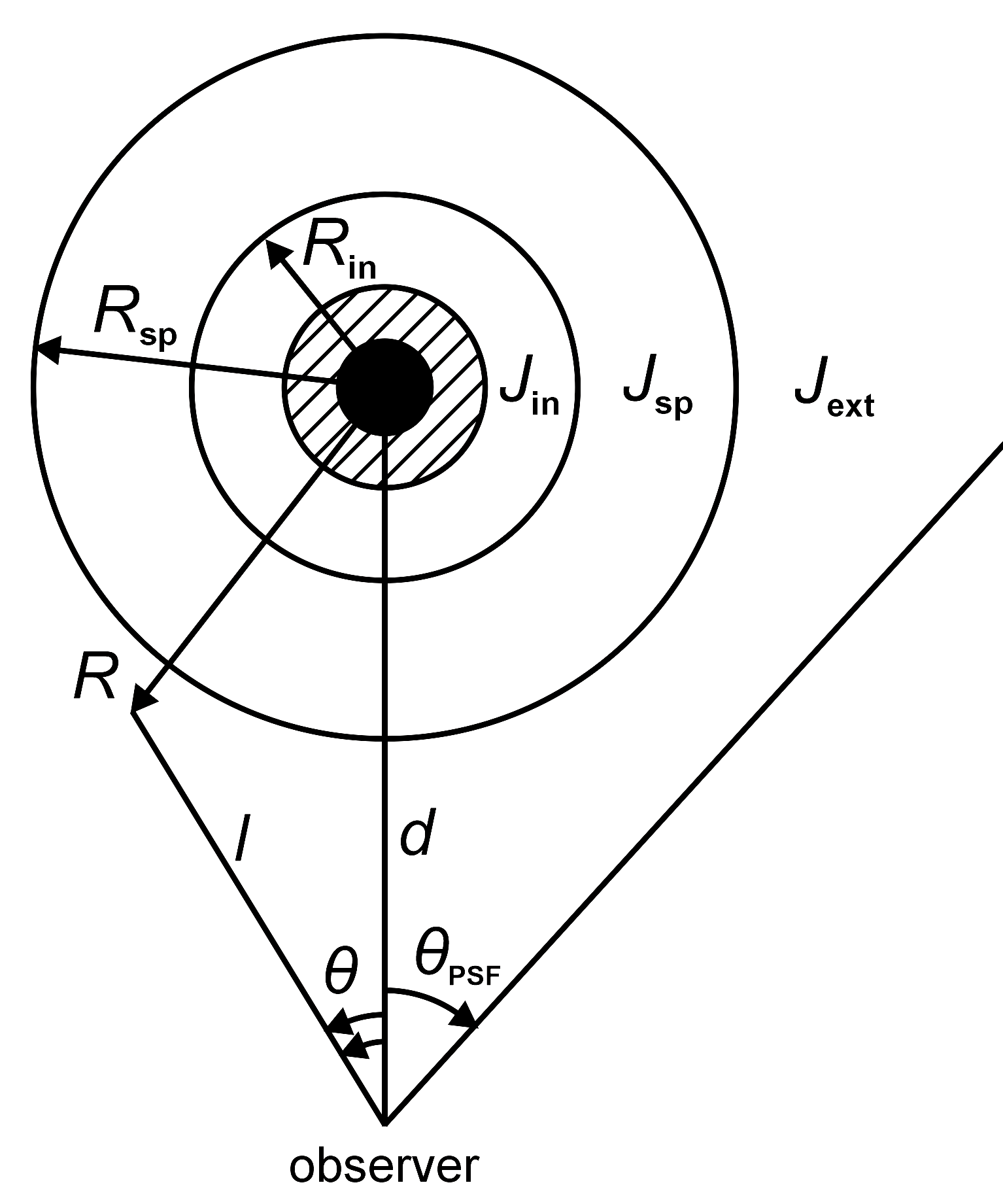}
\caption{\label{fig:1}The scheme of central section of DM density spike: the black disk denotes SMBH, the shaded region has radius 2$R_\bullet$ and effectively null DM density, then the inner spike core and main spike layer follow. Each layer contains respective $J$-factor component symbol.}
\end{figure}

Next let us define DM density distribution. The spike density profile is approximated conventionally by the following piecewise power law:
\begin{equation}\label{eq:rho}
	\rho(R) = \begin{cases}
		0, &  R < 2R_\bullet = 4GM/c^2; \\
		\rho_{in}(R/R_{in})^{-\gamma_{in}}, & 2R_\bullet \leqslant R < R_{in}; \\
		\rho_0(R/R_{sp})^{-\gamma}, & R_{in} \leqslant R < R_{sp}; \\
		\rho_{ext}(R), & R \geqslant R_{sp}.
	\end{cases}
\end{equation}
This profile has four distinct spherical layers, which are sketched in figure \ref{fig:1}. At distances smaller than two Schwarzschild radii (2$R_\bullet$) DM particles do not have stable orbits and fall into SMBH, hence the density is effectively zero. This radial bound was precisely derived in \cite{2013PhRvD..88f3522S}. Meanwhile, some previous studies \cite{2010PhRvD..82h3514G,2015PhRvD..92d3510L,2023PhRvD.108j3042C} used older incorrect bound 4$R_\bullet$, which underestimates the total emission flux from spike. The next layer at $2R_\bullet \leqslant R < R_{in}$ represents possible inner core of the spike, where the density profile is much less steep in comparison with the main spike volume at $R_{in} \leqslant R < R_{sp}$ due to density saturation by too rapid particle annihilation. And beyond the sphere of SMBH gravitational influence at $R \geqslant R_{sp}$, DM density follows the unperturbed global halo profile $\rho_{ext}(R)$. The saturation density $\rho_{in}$ can be estimated from the simple condition, that any particle meets some other particle at least once per SMBH lifetime $t_\bullet$:
\begin{equation}\label{eq:rho_in}
	\rho_{in} = m_x/(\langle\sigma v\rangle t_\bullet).
\end{equation}
The radius of the whole spike corresponds to that of SMBH sphere of influence and can be defined as 
\begin{equation}\label{eq:R_sp}
	R_{sp} = \langle bGM/v^2 \rangle = bGM/\sigma_c^2,
\end{equation}
where $b \sim 1$ (exact values are discussed below). It is easy to show, that for the simple case of isotropic Maxwellian velocity distribution $\langle 1/v^2 \rangle = 1/\sigma_c^2$ with $\sigma_c$ being 1D velocity dispersion. Then assuming the same velocity distribution for DM and baryons, we can identify $\sigma_c$ with the well-observable quantity -- namely, the stellar velocity dispersion along line of sight in a galactic center. $R_{in}$ can be calculated from the natural requirement of density continuity at this bound (see eqs. \eqref{eq:rho}):
\begin{equation}\label{eq:R_in}
	\rho_{in} = \rho_0(R_{in}/R_{sp})^{-\gamma} \rightarrow R_{in} = R_{sp}(\rho_0/\rho_{in})^{1/\gamma}.
\end{equation}
Finally, $\rho_0 = \rho_{ext}(R_{sp})$ comes from the galactic halo density distribution.

At this point we can calculate $J$-factor for eq. \eqref{eq:flux}. As can be seen from figure \ref{fig:1}, $J$-factor contains three components: $J_{in}$ from the spike inner core, $J_{sp}$ from the main spike volume and $J_{ext}$ from DM halo around spike. As was mentioned above, any spike would be contained inside relevant PSF. This allows to integrate $\rho^2(R)$ simply over the whole spike volume for $J$-factor calculation:
\begin{equation}\label{eq:J_sp}
	J_{in}+J_{sp} = \frac{4\pi}{d^2}\int\limits_{2R_\bullet}^{R_{sp}}\rho_{sp}^2(R)R^2dR,
\end{equation} 
where $d$ is the distance to SMBH, which satisfies $d\ggg R_{sp}$. The gamma-ray flux from halo outside the spike is not the subject of our main interest. However, it must be included (at least approximately) for correctness and completeness of our model. Thus, $J_{ext}$ is calculated as the difference of $J$-factors inside PSF and $R_{sp}$ for the unperturbed halo profile (see figure \ref{fig:1} again): 
\begin{equation}\label{eq:J_ext}
	J_{ext} = \int\limits_{V_\text{PSF}}\rho_{ext}^2(R)\frac{dV}{l^2} - \frac{4\pi}{d^2}\int\limits_0^{R_{sp}}\rho_{ext}^2(R)R^2dR,\quad R^2=l^2+d^2-2ld\cos\theta,
\end{equation}
where $V_\text{PSF}$ denotes the halo volume contained inside PSF. Combining eqs. \eqref{eq:rho}, \eqref{eq:J_sp}, \eqref{eq:J_ext} and performing simple analytical integration, we can write out the final expression for $J$-factor:
\begin{equation}\label{eq:J}
	\begin{split}
		& J(\langle\sigma v\rangle, m_x, M, t_\bullet, \sigma_c, d, \rho_0, \gamma_{in}, \gamma, b) = J_{in}+J_{sp}+J_{ext}; \\
		& J_{in} = \begin{cases}
			0, &R_{in}\leqslant 2R_\bullet; \\
			\frac{4\pi\rho_{in}^2}{d^2(3-2\gamma_{in})}\left(R_{in}^3-8R_\bullet^3\left(\frac{R_{in}}{2R_\bullet}\right)^{2\gamma_{in}}\right), &R_{in}>2R_\bullet,~ \gamma_{in} \neq 3/2;
		\end{cases}	\\
		& J_{sp} = \begin{cases}
			\frac{4\pi\rho_0^2}{d^2(3-2\gamma)}\left(R_{sp}^3-\max(2R_\bullet,R_{in})^3\left(\frac{R_{sp}}{\max(2R_\bullet,R_{in})}\right)^{2\gamma}\right), &\gamma \neq 3/2; \\
			4\pi\rho_0^2 R_{sp}^3\ln(R_{sp}/\max(2R_\bullet,R_{in}))/d^2, &\gamma=3/2;
		\end{cases} \\
		& J_{ext} = 2\pi\int\limits_{l_\text{min}}^{l_\text{max}}dl\int\limits_0^{\theta_\text{PSF}}\rho_{ext}^2\left(\sqrt{l^2+d^2-2ld\cos\theta}\right)\sin\theta d\theta - \frac{4\pi}{d^2}\int\limits_0^{R_{sp}}\rho_{ext}^2(R)R^2dR.
	\end{split}
\end{equation}
The first eq. in \eqref{eq:J} lists explicitly all important variables, which defines $J$ and the gamma-ray flux. Here we have exceptional situation, when $J$-factor depends not only on macroscopic properties of DM halo, but also on microscopic parameters of DM particles. This is conditioned by eq. \eqref{eq:rho_in} and makes the flux dependence on annihilation cross section in eq. \eqref{eq:flux} non-linear, at least when the inner spike core exists. The latter is not always true, since DM density in the spike may be below the saturation density \eqref{eq:rho_in} everywhere. In this case $R_{in}\leqslant 2R_\bullet, J_{in}=0$ and the whole spike is homogeneous with $\rho_{sp}\propto R^{-\gamma}$. 

\subsection{\label{sec:par}Choice of the spike parameter values}
Now let us discuss the model parameter values and their uncertainties. PSF angular radius was chosen to be $\theta_\text{PSF}$ = 0.06\degree. In general, this quantity depends on photon energy indeed. However, this energy dependence was neglected, since it does not affect the calculation of spike signal and defines only the flux from external halo, which does not play a big role for our objectives. The chosen constant PSF size reflects CTA angular resolution in the middle of relevant energy range at $E \sim 1$ TeV \cite{CTA}.

SMBH age was set to be $t_\bullet = 10^{10}$ years. The inner spike core was studied in \cite{2007PhRvD..76j3532V} and \cite{2016PhRvD..93l3510S}. These two works obtained (independently) the same value for the profile slope $\gamma_{in}=1/2$ for s-wave annihilation. I took this value as fixed and quite certain. The biggest model uncertainties come from three parameters: the density slope $\gamma$, spike "base" density $\rho_0$ and size parameter $b$. Next paragraphs elaborate in details each of these parameters.

The spike density profile was derived for the first time in \cite{1999PhRvL..83.1719G} in the frame of highly idealized model of adiabatic SMBH growth without any disturbance of spike by stars and other entities. This model yielded quite universal spike density profile: for the wide range of initial unperturbed halo density slopes $0\leqslant \gamma_0 \leqslant 2$, the final slope resides in the narrow range $2.25\leqslant \gamma \leqslant 2.5$. Recently \cite{2024MNRAS.527.3196S} derived important generalization: the adiabatic spike slope for Einasto halo profile, which has variable slope $\gamma_0(R)$. Such spike appeared to be quite similar having $\gamma\approx 2.3$ for relevant Einasto halo profiles. The opposite scenario is instantaneous SMBH formation, which results in a very shallow spike with $\gamma=4/3$ \cite{2001PhRvD..64d3504U}. Omitting exotic scenarios with off-centered or binary \cite{2002PhRvL..88s1301M} SMBHs, we can consider the adiabatic and instantaneous formation scenarios as the most optimistic and pessimistic cases for our model. A realistic scenario is likely some mixture of those extremes. And another important effect is the spike dynamical evolution due to interaction with a dense stellar environment around galactic center. This effect was modeled in e.g. \cite{2004PhRvL..93f1302G,2022PhRvD.106d3018S}, which firmly derived the resulting spike slope $\gamma=3/2$. Thus, theoretical models have significant uncertainties in prediction of spike structure. On observational side, it is also very challenging to infer the former. In this context \cite{2024MNRAS.527.3196S} employed very precise observations of so-called S-stars, which are the closest to MW*, in order to derive constraints on spike properties. However, the sensitivity of this method was not sufficient to fully exclude even the densest adiabatic spike. Later \cite{2025arXiv250610122S} developed a novel interesting method to determine the spike profile using reverberation mapping measurements of active galactic nuclei. Although the majority of studied objects did not demonstrate any conclusive inferences too, one very remote SMBH (3C 390 at $z=0.30$) yielded $\gamma\approx 1.5$ with tangible significance. Having this "global picture", I decided to choose $\gamma=4/3\div2.3$ as the continuous range of possible values and to not stick to particular values only. The boundary values generate the smallest and largest emission fluxes from the spike. I designate these cases by MIN and MAX notation traditionally. But the medium expectation case (MED), i.e. the most realistic case, seems to have $\gamma\approx 1.5$ rather than the formal mean of limiting values $\langle\gamma\rangle\approx 1.8$.

Next let us discuss the halo density $\rho_0$ at the external edge of spike. As was already mentioned, it is highly uncertain too, because both theoretical simulations (e.g. \cite{2016MNRAS.456.3542T}) and observations of stellar kinematics (e.g. \cite{2012A&A...546A...4T}) allow both cuspy and cored halo density profiles, which have respectively high and low $\rho_0$ values and, hence, would provide bright and faint emissions from spike. Thus, cored and cuspy profiles determine MIN--MAX range of possibilities. And I defined MED profile as the geometric average of MIN and MAX profiles, which is more representative here in comparison with usual arithmetic average, since possible density values may span more than two orders of magnitude. Specific profile choices are described for each relevant galaxy individually in the next section.

Finally, we have to evaluate the possible spike radius values. Just from a general intuition, this radius should be about the radius of SMBH sphere of influence $R_h$ and can be estimated as $GM/\sigma_c^2$, which reflects the radius of circular orbit for DM particles in a galactic center. The exact radius has certain model dependence, which is encoded by the prefactor $b$. The majority of spike studies cited above employed the value $b=0.2$. This value originated probably from \cite{2004cbhg.symp..263M} (eq. (10) there). However, the author considered only stellar (not DM) spike in this work and used his own peculiar definition for the radius of SMBH sphere of influence ($r_h$ there): $M_*(R<R_h)=2M$, i.e. the stellar mass inside this radius is equal to two SMBH masses. This definition implies a non-trivial expression for $R_h$ -- \cite[eq. (4)]{2004cbhg.symp..263M}, which matches the simple expression $R_h=GM/\sigma_c^2$ only for the particular case of isothermal stellar density profile. However, in reality the former can be and likely is much less steep than isothermal. This implies $R_h>GM/\sigma_c^2$ and, hence, $b>0.2$ in my notation. Therefore, $b=0.2$ tends to be the minimal possible value. At the same time, the detailed observational study of MW nuclear star cluster \cite{2018A&A...609A..27S} firmly determined $R_h=(3.1\pm0.3)$ pc using Merritt's definition of $R_h$. Multiplication of this value by 0.2 yields DM spike radius, which corresponds to $b\approx 0.6$, if to substitute MW* parameters into my definition \eqref{eq:R_sp}. Saying in other words, $0.2R_h$ in Merritt's notation corresponds to $0.6R_h$ in the notation used by e.g. \cite{2014PhRvL.113o1302F,2015PhRvL.115w1302S,2020PhRvD.102b3030C,2023JCAP...08..063B}, if to rely on \cite{2018A&A...609A..27S}. Considering the latter observational justification of $b=0.6$ value and also its proximity to intuitive $b=1$ value, I set $b=0.6$ as the most realistic (MED) value. And several mentioned studies may had underestimated significantly the emission flux from spikes using $b=0.2$. Some other previous works \cite{2015PhRvD..92d3510L,2023PhRvD.108j3042C} employed the different method to obtain the upper limit on $R_{sp}$, specifically for M87*. This limit was derived from the requirement, that DM mass inside $R_h$ must not exceed the uncertainty of kinematical measurements of SMBH mass, yielding $b\approx 1.6$. However, SMBH mass uncertainty likely has other large contributions besides the systematic one, which is caused by DM "alteration" of stellar motions. Thus, such $b$ value tends to an opposite overestimating extreme. In this situation I chose modest and neat value $b=1.0$ for MAX case, which also naturally places MED value 0.6 to the middle of MIN--MAX range $b=0.2\div1.0$.

\section{\label{sec:all}Comparison of nearby SMBHs}

At this point we have outlined the general formalism completely and can proceed to analysis of specific targets. At the first step, I constructed the short list of relevant nearby ($d\leqslant100$ Mpc) SMBHs based on their mass and distance using a general catalog (e.g. \cite{SMBH}). The mass plays even more important role than the distance, since the expected emission flux scales as $\Phi\propto d^{-2}M^3$ according to eqs. \eqref{eq:flux},\eqref{eq:J}. The obtained short list is presented in table \ref{tab}, which displays all relevant data for each object.

\begin{figure}[h]
	\centering
	\includegraphics[width=0.495\textwidth]{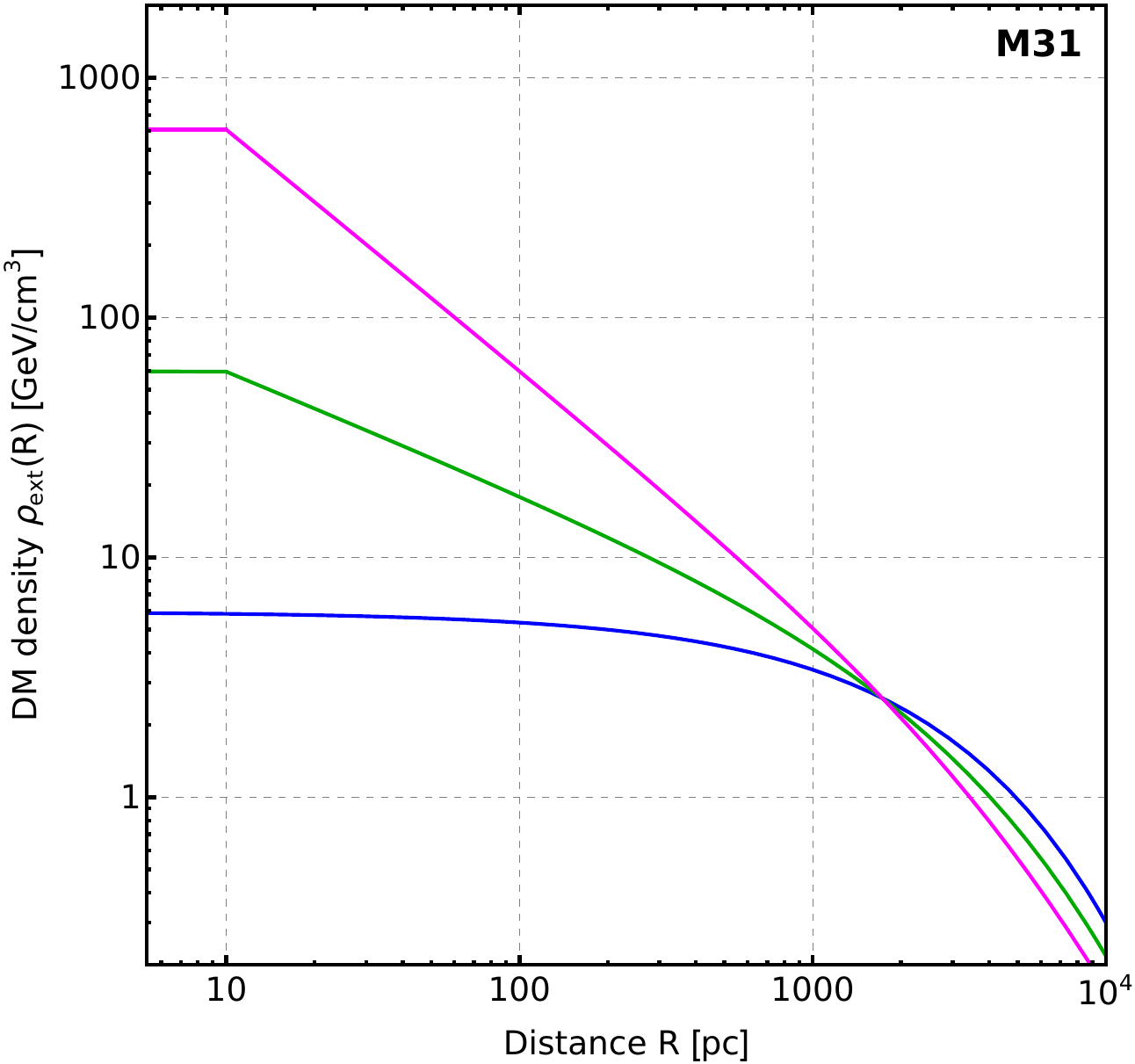}
	\hfill
	\includegraphics[width=0.495\textwidth]{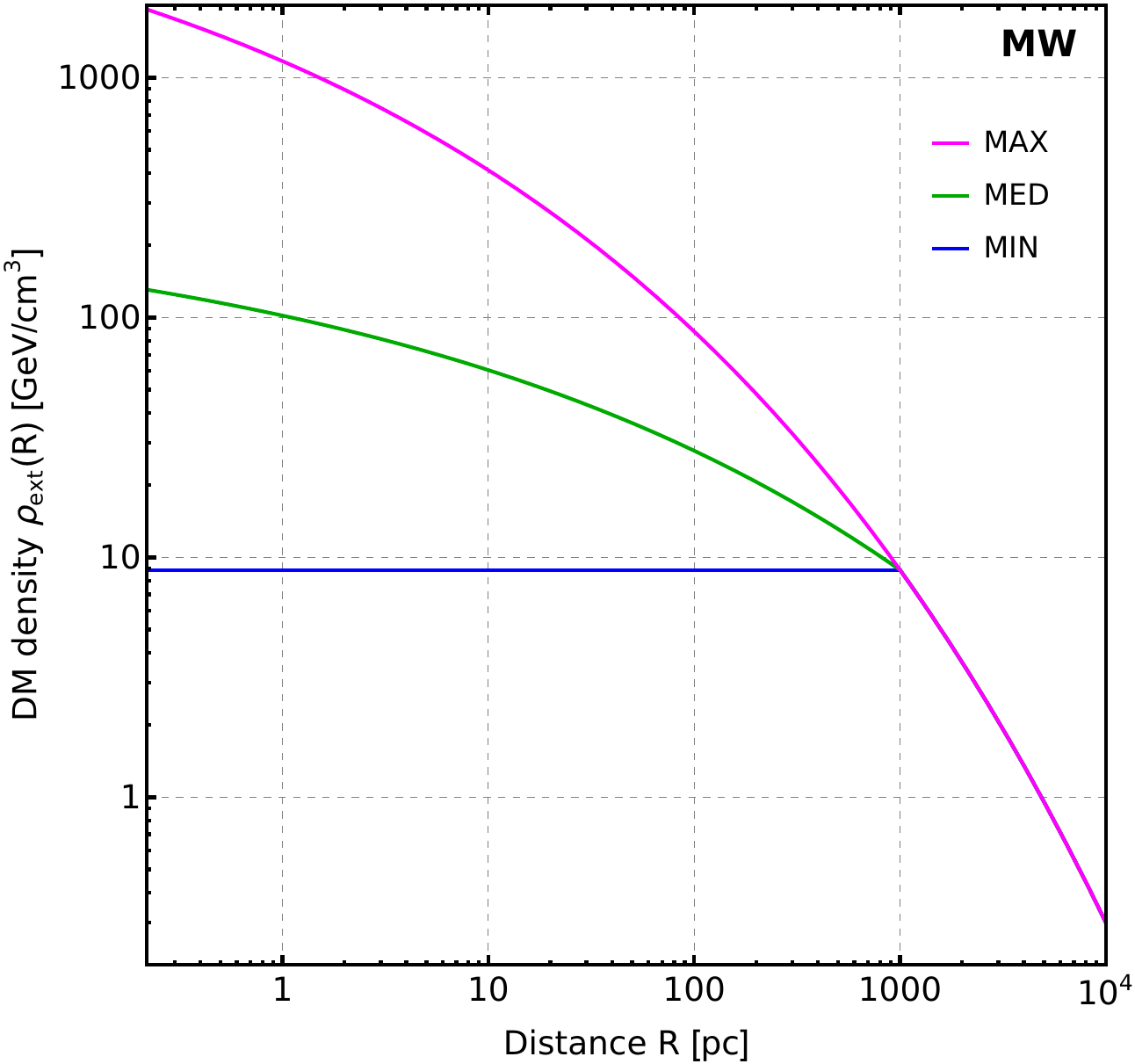}
	\caption{\label{fig:rho}DM galactic halo density profiles for M31 and MW for three accepted model configurations. The distance axis origins correspond to the smallest possible spike radii $R_{sp}(b=0.2)$.}
\end{figure}

I estimated $J$-factor as the main signal predictor for each SMBH using MED model configuration described above. It implies relatively modest spike densities, and the annihilation core does not develop for TeV-scale WIMPs, i.e. $J_{in}=0$. $J_{ext}$ was also omitted in these estimations, since the pure spike brightness is of interest here. $\rho_0$ estimation, which is required for $J_{sp}$ calculation, was quite challenging due to large uncertainties. The limiting (i.e. MAX and MIN) DM halo density profiles for MW were taken from \cite{2021JCAP...01..057A} for equivalence of comparison with my results. Those are Einasto profiles: usual and with 1 kpc flat density core, respectively. MED profile was derived as the geometric average of MIN and MAX, as was mentioned above. All three profiles are shown in figure \ref{fig:rho}. Then $\rho_0$ was calculated as $\rho^\text{MED}(R_{sp}^\text{MED})$. For M31 I employed the profiles, which were adopted and justified in my previous work \cite[section IIIB]{2022PhRvD.106b3023E}: mild Einasto and NFW, also shown in figure \ref{fig:rho} together with their average. Regarding other galaxies, unless the explicit density profiles were found in literature (like it was for e.g. M87), the following procedure was applied. DM halo mass and concentration were found in the cited papers for each galaxy. Then I converted these data into the density profile parameters using simple relations (e.g. \cite[eq. (14)]{2013PhRvD..88b3504E}). The basic two-parametric density profiles were used: Burkert as MIN and NFW as MAX. The former is defined as $\rho(R)=\rho_s/((1+R/R_s)(1+(R/R_s)^2))$ \cite{Burkert} and has the following relation between virial halo concentration $c_v$ and radius $R_v$: $R_v\simeq1.521c_vR_s$ (for NFW it is simply $R_v=c_vR_s$). The obtained $\rho_0$ values anticorrelate well with the galactic halo mass in accordance with basic $\Lambda$CDM paradigm: smaller halos have higher densities. Thus, for example, the giant M87 halo has the mass $M_v\sim10^{14}M_\odot$ and tiny central density $\rho_0\approx0.52$ GeV/cm$^3$ (although, this mass value is still debatable \cite{2014A&A...570A.132S}). The central velocity dispersion $\sigma_c$ correlates well with SMBH mass $M$ in accordance with the famous relation. M87 is again an outlier, but this may be just due to exceptionally high quality of recent observations by JWST \cite{2025arXiv251002439A}.

Looking at the obtained $J_{sp}$ values in table \ref{tab}, we see that MW* and M31* are distinct leaders, and all other objects can be discarded from observational program (at least in the context of s-wave annihilating WIMPs). Paradoxically M87* appears to have the smallest $J_{sp}$! This is caused substantially by very low central halo density $\rho_0$. Therefore, a potential of this object was overestimated in the previous studies, which was also noted in \cite{2025PhRvD.111k5033P}. Interestingly, moderate SMBH NGC 3115* demonstrates exactly same $J_{sp}$ as huge NGC 3842*, which is located by 10 times further. Indeed the spike model has large uncertainties, but they influence the absolute $J_{sp}$ values and not much the SMBH ranking. Considering MW* and M31*, they manifest very close values $J_{sp}^\text{MW}\approx3J_{sp}^\text{M31}$. Therefore, it is very not obvious, which of these two targets would provide better sensitivity, and M31* was unfairly missed in previous studies. Thus, the optimal observational targets have been identified -- let us research them in details.

\begin{table}[t]
\centering
\begin{tabular}{|c|c|c|c|c|c|}
\hline
\rule{0pt}{1em}
SMBH host & $d$ [Mpc]{\cite{NED}} & $M~[M_\odot]$ & $\sigma_c$ [km/s]{\cite{HyperLeda}} & $\rho_0$ [GeV/cm$^3$] & $J_{sp}$ [GeV$^2$/cm$^5$] \\
\hline
\rule{-4pt}{1em}
MW & 0.0083 {\cite{2022A&A...657L..12G}} & $4.3\cdot 10^6$ {\cite{2022A&A...657L..12G}} & 130 {\cite{2018A&A...616A..83V}} & 110 {\cite{2021JCAP...01..057A}} & $2.6\cdot 10^{16}$ \\
M31 & 0.76 {\cite{2021ApJ...920...84L}} & $1.4\cdot10^8$ {\cite{2005ApJ...631..280B}} & 150 & 47 {\cite{2022PhRvD.106b3023E}} & $8.2\cdot 10^{15}$ \\
NGC 3115 & 10 & $9.6\cdot10^8$ {\cite{2009ApJ...698..198G}} & 260 & 27 {\cite{2017MNRAS.468.3949A,2014A&A...570A.132S,2022AJ....163..234K}} & $1.7\cdot 10^{14}$ \\
M104 & 11 & $1.0\cdot10^9$ {\cite{2024ApJ...965..128Y}} & 230 & 13 {\cite{2011ApJ...739...21J}} & $7.8\cdot 10^{13}$ \\
M60 & 17 & $2.1\cdot10^9$ {\cite{2009ApJ...698..198G}} & 330 & 11 {\cite{2017MNRAS.468.3949A}} & $2.3\cdot 10^{13}$ \\
M87 & 17 & $6.5\cdot10^9$ {\cite{2019ApJ...875L...6E}} & 450 {\cite{2025arXiv251002439A}} & 0.52 {\cite{2011ApJ...729..129M,2022ApJ...929...17D}} & $2.3\cdot 10^{11}$ \\
M84 & 17 & $1.5\cdot10^9$ {\cite{2009ApJ...698..198G}} & 280 & 6.8 {\cite{2017MNRAS.468.3949A}} & $9.0\cdot 10^{12}$ \\
NGC 4889 & 97 & $2.0\cdot10^{10}$ {\cite{2012ApJ...756..179M}} & 390 & 4.7 {\cite{2007MNRAS.382..657T}} & $4.0\cdot 10^{13}$ \\
NGC 3842 & 100 & $1.0\cdot10^{10}$ {\cite{2012ApJ...756..179M}} & 310 & 14 {\cite{2012ApJ...756..179M}} & $1.7\cdot 10^{14}$ \\     		
\hline
\end{tabular}
\caption{\label{tab}The list of potentially promising nearby and massive SMBHs arranged by distance. The second column contains distance, the third -- mass, the fourth -- stellar velocity dispersion in the center of host galaxy, the fifth -- employed DM halo density at the spike border and the sixth -- calculated spike $J$-factor. The origin of all data is provided (the first row contains default data source). MED model parameter values were used in calculation: $\gamma=1.5,~b=0.6$, medium density $\rho_0$.}
\end{table}

\section{\label{sec:cta}CTA sensitivity to WIMP signal near M31* and MW*}

\begin{figure}[t]
	\centering
	\includegraphics[width=0.65\textwidth]{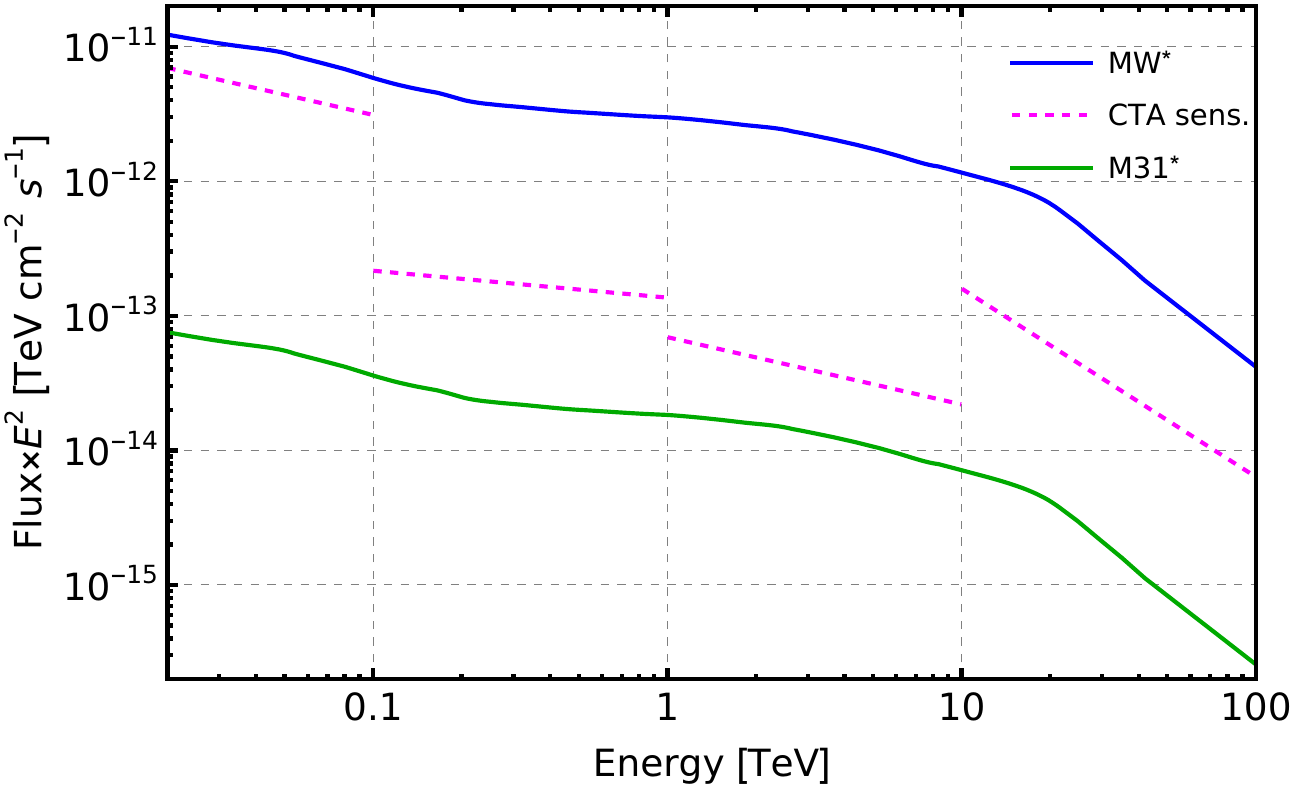}
	\caption{\label{fig:flux}The blue line shows the spectrum of J1745--290 source taken from \cite[figure 3]{2021ApJ...913..115A} and extrapolated to 100 TeV by power law. The green line shows the same spectrum assigned to M31 center by rescaling to M31 gamma-ray flux measured at $\approx$ 0.01 TeV. The dashed line shows approximate CTA sensitivity to a point source with relevant power-law spectrum in the chosen wide energy bins for $T=100$ h. More details are in section \ref{sec:cta}.}
\end{figure}

MW* DM density spike received probably the biggest attention due to proximity. But this object is somewhat challenging from observational point of view: MW* region is very bright in gamma rays, and it is difficult to separate various emission sources there and isolate small spike unambiguously. The bright gamma-ray source J1745--290 \emph{potentially} corresponds to MW*, its spectrum was measured well at all energies up to $\approx$ 40 TeV \cite{2016Natur.531..476H,2021ApJ...913..115A} (shown with extrapolation in figure \ref{fig:flux}). However, these studies acknowledge, that currently available imaging capabilities are not sufficient to detect MW* confidently. J1745--290 contains likely several sources: SMBH accretion disk, nearby pulsar wind nebula, dense gas shining due to CRs and potentially DM annihilation spike, which could have just a minor contribution. CTA will have better capabilities and indeed good chances to disentangle this complicated source. But an exact outcome is unpredictable apriori, i.e. without having real data. In unlucky case CTA may see the same mixed point source and be able just to extend its spectrum to higher energies. I estimated the sensitivity to annihilating WIMPs in MW* spike in the frame of this conservative assumption and the simple requirement for the emission flux from spike to not exceed the measured spectrum. These two postulates are very crude indeed: all the observed emission is artificially associated with DM in this approach. But it is difficult to construct better method now, partly due to impossibility to separate contributions from various emission sources mentioned above. Although such sensitivity estimate is very conservative, at the same time it is very \emph{model-independent} and will provide at least some indicative comparison with M31* case. My procedure for MW* majorly repeats the constraints derivation in \cite{2023JCAP...08..063B}. But both targets must be analyzed in the frame of same spike model, hence it would not be useful to employ the results, which are based on slightly different model.

Now let us consider M31*, which is the main subject of interest in my work. As was mentioned in section \ref{sec:i}, M31 nucleus is very quite in non-thermal emissions. So far only flux upper limits were obtained at the energies above $\approx0.01$ TeV \cite{2020ApJ...893...16A}. Therefore, potential DM spike signal is much less contaminated by background emissions. At the same time, as was found out above, M31* $J$-factor is comparable to that for MW*, which provides better expectations for "signal-to-noise" ratio. CTA may either detect or not M31 regular astrophysical emissions. The case of non-detection will be probably the best for DM search. Let us check this possibility assuming the same spectrum for M31 central astrophysical (i.e. not related to DM) emission as J1745--290 has, but renormalized according to the absolute spectral flux measured by Fermi-LAT from M31 central region at 10 GeV \cite{2023ApJ...945L..22X}. Such rescaled spectrum is shown by the green line in figure \ref{fig:flux} and likely overestimates the expected flux for CTA, since its PSF is smaller than that of Fermi-LAT. Nevertheless, let us evaluate whether CTA will achieve detection of such hypothetical M31 central source, assuming it to be point-like.

CTA sensitivity to a point source is defined mainly by photon flux from it and the background rate, which comes from contaminating residual CRs and is isotropic over sky. A precise calculation of sensitivity is a complicated procedure, which employs instrument response functions and is usually performed by the collaboration. I adopted the following simplified methodology in order to obtain reasonable estimates. Three criteria for source detection were required (in conjunction) in accordance with CTA performance evaluation by the collaboration \cite{CTA}: 5$\sigma_s$ statistical significance, at least 10 signal photons and a signal/background ratio of at least 1/20. I.e. in respective order:
\begin{equation}\label{eq:DR}
	\left\{ \begin{array}{l}
		S\geqslant5, \\
		N_s\geqslant10, \\
		N_s/N_b\geqslant0.05.
	\end{array} \right.
\end{equation}
The following equation for statistical significance was employed, which is conventionally used for IACTs \cite[eq. (17)]{1983ApJ...272..317L}:
\begin{equation}\label{eq:S}
	S^2/2 = (N_s+N_b)\ln\frac{(1+\alpha)(N_s+N_b)}{\alpha(N_s+N_b)+N_b}+\dfrac{N_b}{\alpha}\ln\frac{(1+\alpha)N_b}{\alpha(N_s+N_b)+N_b},
\end{equation}
where $\alpha$ is the ratio of on-source to off-source exposure times and set to be $\alpha=0.2$ according to \cite{2013APh....43..171B}. Photon numbers were calculated as
\begin{align}
	\label{eq:Nb}
	N_b & = \int\limits_{E_1}^{E_2} B(E)T\Omega_\text{PSF}(E)dE = 2\pi T\int\limits_{E_1}^{E_2} B(E)T(1-\cos\theta_\text{PSF}(E))dE,   \\
	\label{eq:Ns}
	N_s & = 0.68\int\limits_{E_1}^{E_2}dE'\int s(E)\frac{1}{\sqrt{2\pi}\sigma_E(E)}\exp\left(-\frac{(E-E')^2}{2\sigma_E^2(E)}\right)TA(E)dE,
\end{align}
where $E_1$ and $E_2$ are the bounds of energy range/bin of interest, $B(E)$ is the instrumental background rate due to CRs, $T$  is the exposure time, $\theta_\text{PSF}(E)$ is the angular resolution (here it has proper energy dependence), prefactor 0.68 reflects photon fraction containment inside PSF, $s(E)$ is the source spectrum, $A(E)$ is the telescope effective area and the exponent models finite energy resolution $\sigma_E(E)$ through Gaussian energy dispersion. I took all the performance data from \cite{CTA} and chose CTA Northern Array, because M31 has declination $\delta\approx+41\degree$ on the sky, hence can be observed mainly from northern site. In order to test whether my simplified methodology provides reasonable precision, I reproduced the differential sensitivity curve provided by CTA collaboration for the case $T=50$ h. My algorithm yielded slightly elevated sensitivity expectedly: the ratio of precise sensitivity flux to my values is in the range 1.2--1.6 for all energy bins, which are equivalent indeed (five bins per decade). This is modest uncertainty comparing with other model systematics. This uncertainty needs to be just kept in mind in interpretation of final results.

Having this algorithm, I estimated CTA sensitivity in the broad decade-wide energy bins to a source, which has power-law spectrum $s(E)\propto E^\beta$ close to J1745--290 spectrum in each of these bins. The exposure time is set to $T=100$ h hereafter, which is approximate anticipated time for M31 observations according to \cite{2023JCAP...08..073M}. The obtained sensitivity is shown in figure \ref{fig:flux} by the dashed line as the power-law spectrum fragments, which are at the threshold of detectability in each bin. The whole photon energy range, which I used in my work, spans from 0.02 TeV to 100 TeV. The lower bound is conditioned by that of CTA operation range and the upper bound corresponds to the highest possible photon energy for the heaviest WIMP. We can see that CTA will detect easily J1745--290 over the whole cited range. And M31 center is below the detection threshold at all the energies. Moreover, this would be likely true for any reasonable spectrum shape -- not only for the trial shape considered here. This confirms the initial guess, that CTA will be able to realize its constraining power \emph{in full} for DM in M31*, since usual nuisance astrophysical emissions should not achieve visibility, and the sensitivity to potential WIMP emission will be mainly limited by the telescope sensitivity only. The above reasoning has the implicit assumption, that the emission spectrum of MW central source is generated mainly by conventional sources, not by DM. But this is a quite realistic scenario.

\begin{figure}[t]
	\centering
	\includegraphics[width=0.495\textwidth]{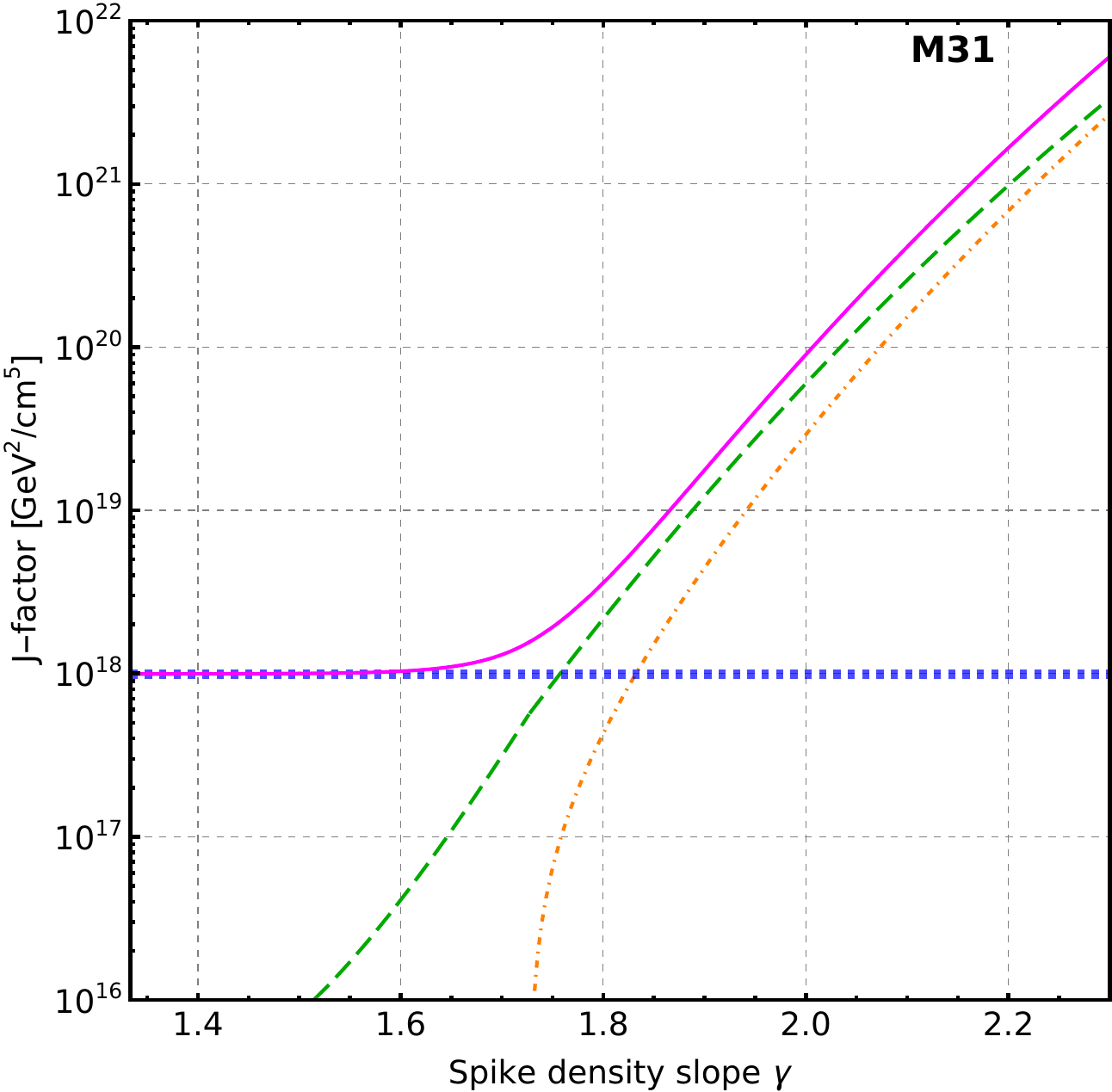}
	\hfill
	\includegraphics[width=0.495\textwidth]{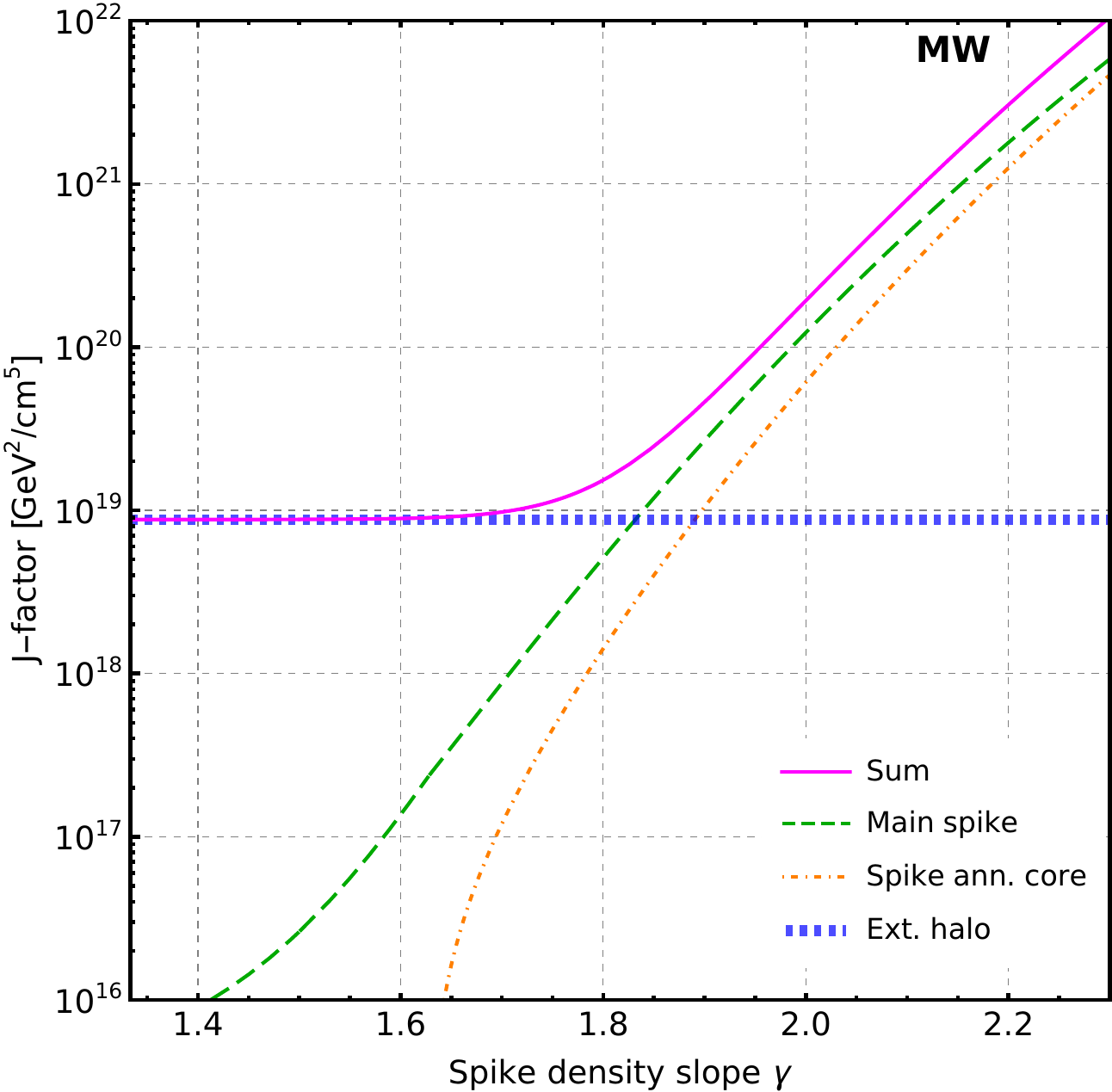}
	\caption{\label{fig:J}The dependencies of various $J$-factor components on the main spike density slope $\gamma$ for both galaxies. MED DM halo density profiles were employed, $b=0.6$, $m_x=3$ TeV, $\langle\sigma v\rangle = 2.1\cdot10^{-26}$ cm$^3$/s (thermal relic). The blue line represents $J_{ext}$, orange line -- $J_{in}(\gamma)$, green line -- $J_{sp}(\gamma)$ and continuous line -- total $J(\gamma)$.}
\end{figure}

Now we can move to derivation of CTA sensitivity to annihilating WIMPs in M31* DM spike, assuming essentially the absence of astrophysical backgrounds. As an intermediate step, let us explore in details the dependence of all $J$-factor components on the key spike model parameter $\gamma$. These dependencies were calculated by eqs. \eqref{eq:J} and are presented in figure \ref{fig:J} for M31 and MW. Other spike model parameters have MED values (written out in table \ref{tab:2}), the trial thermal WIMP with $m_x=3$ TeV was chosen. We can notice very close similarity between the spike $J$-factors of two objects, as was envisioned before. Regarding $J_{ext}$ value, it is indeed very uncertain due to both DM halo profile uncertainties and also $\theta_\text{PSF}$ ambiguity. But the certain medium value was chosen in order to assess the ratio of various $J$ components at least roughly. Thus, we can see that seemingly most realistic spike with $\gamma=1.5$ (discussed in section \ref{sec:par}) would be much below the diffuse DM halo emission, likely for any possible $J_{ext}$ values. This is an unfortunate finding. The spike may add any tangible progress to WIMP constraints, which come from observations of diffuse DM halo, only if $\gamma\gtrsim1.7-1.8$, which reflects rather optimistic case of so-called Bahcall-Wolf density cusp \cite{2006ApJ...648..890M}. And the spike inner core due to annihilation emerges at the same $\gamma$ values. The ratio $J_{in}/J_{sp}$ increases with $\gamma$, but never exceeds 1. But other WIMP masses would imply other ratios: lighter WIMPs have lower $\rho_{in}$ according to eq. \eqref{eq:rho_in}, therefore, the spike inner core would be larger; and vice versa for heavier WIMPs. Overall, the spike $J$-factor has very steep dependence on $\gamma$.
\begin{table}[t]
	\centering
	\begin{tabular}{|c|c|c|c|}
		\hline
		Model parameter & MIN & MED & MAX \\
		\hline
		$b$ & 0.2 & 0.6 & 1.0 \\
		\rule{-4pt}{1em}
		$R_{sp}^\text{M31}$ [pc] & 5.3 & 16 & 27 \\
		\rule{-4pt}{1em}
		$\rho_0^\text{M31}$ [GeV/cm$^3$] & 5.9 & 47 & 230 \\
		\rule{-4pt}{1em}
		$R_{sp}^\text{MW}$ [pc] & 0.22 & 0.66 & 1.1 \\
		\rule{-4pt}{1em}
		$\rho_0^\text{MW}$ [GeV/cm$^3$] & 8.7 & 110 & 1100 \\     		
		\hline
	\end{tabular}
	\caption{\label{tab:2}DM density spike model parameter values for both galaxies for three accepted configurations.}
\end{table}

I estimated CTA sensitivity for WIMP mass range $m_x=(0.1-100)$ TeV. More specifically, the limiting annihilation cross sections were computed individually for 16 discrete mass values in this range, and then the obtained points were machine-interpolated on the parameter plane. Two traditional representative annihilation channels were computed: $\chi\chi\rightarrow b\bar{b}$ and $\chi\chi\rightarrow \tau^+\tau^-$. The limits for any other "pure" channel or an arbitrary mixture are likely enclosed between the limits for chosen channels, as was demonstrated in e.g. \cite[figure 2]{2023ecrs.confE.120E}.

The whole procedure here contained two major stages. DM signal spectrum shape depends on $m_x$ and annihilation channel only. Taking this into account, at the first step I took the spectra for each mass $s(E,m_x)=F(m_x)dN_\gamma/dE(E,m_x)$, substituted them into eqs. \eqref{eq:S}--\eqref{eq:Ns} and solved them semi-numerically finding the following for each $m_x$: the optimal energy range $[E_1,E_2]$, which provides the \emph{minimal} possible detectable amplitude $F(m_x)$ satisfying the requirements \eqref{eq:DR}, and this minimal amplitude itself. At the second stage I substituted the obtained limiting spectra in \eqref{eq:flux}, i.e. put $d\Phi/dE=s(E,m_x)$, which yields
\begin{equation}\label{eq:F}
\frac{\langle\sigma v\rangle}{8\pi m_x^2}J(\langle\sigma v\rangle, m_x, M, t_\bullet, \sigma_c, d, \rho_0, \gamma_{in}, \gamma, b) = F(m_x).
\end{equation}
And the limiting cross section $\langle\sigma v\rangle$ was computed numerically from this equation for each $m_x$, annihilation channel and parameter configuration of interest; substituting also eqs. \eqref{eq:rho_in}--\eqref{eq:R_in} and \eqref{eq:J}. CTA has a poor sensitivity to light WIMPs due to the limits of operational energy range. For this reason, I naturally decided to add the current Fermi-LAT limits on the flux from M31 center \cite{2023ApJ...945L..22X} to the derived CTA sensitivity. I.e., the requirement $d\Phi/dE\leqslant$ [Fermi-LAT upper limit] was added to eq. \eqref{eq:F}, and the smallest $\langle\sigma v\rangle$ from both equations was chosen for each mass. This completed the derivation of such combined sensitivity to WIMPs in M31* spike. Regarding MW*, the indicative constraints were derived from the simple condition $d\Phi/dE=$ [J1745--290 flux upper limit from \cite[figure 3]{2021ApJ...913..115A}]. Although such constraints are rather weak, as was explained above, they also serve an important purpose to check overall correctness of my procedure by comparison with similar constraints obtained independently in \cite{2023JCAP...08..063B}.

\begin{figure}[t]
	\centering
	\includegraphics[width=0.495\textwidth]{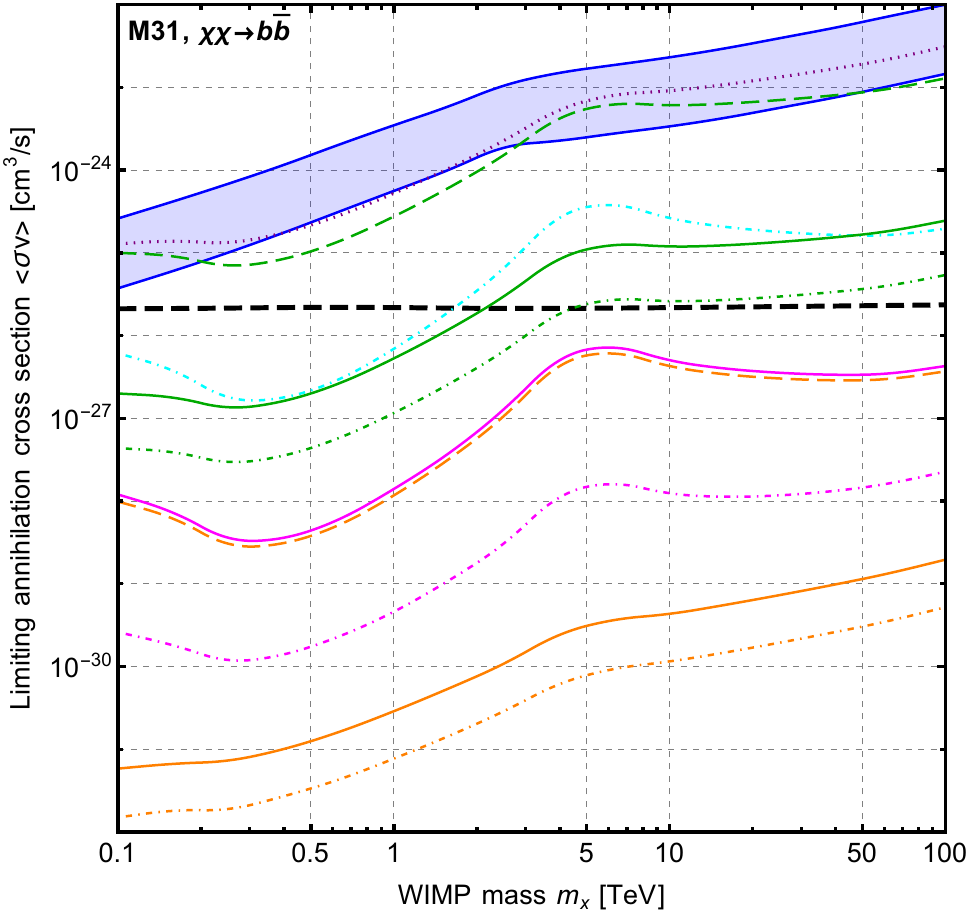}
	\hfill
	\includegraphics[width=0.495\textwidth]{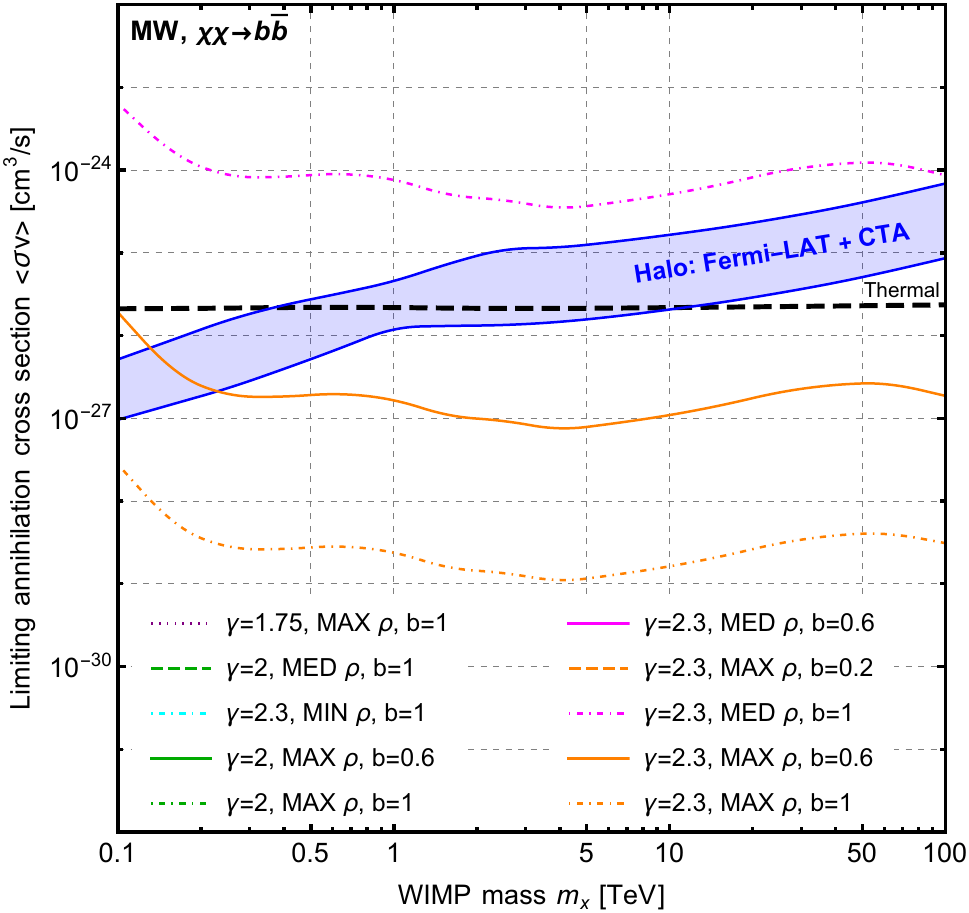}
	
	\vspace{10pt}
	\includegraphics[width=0.495\textwidth]{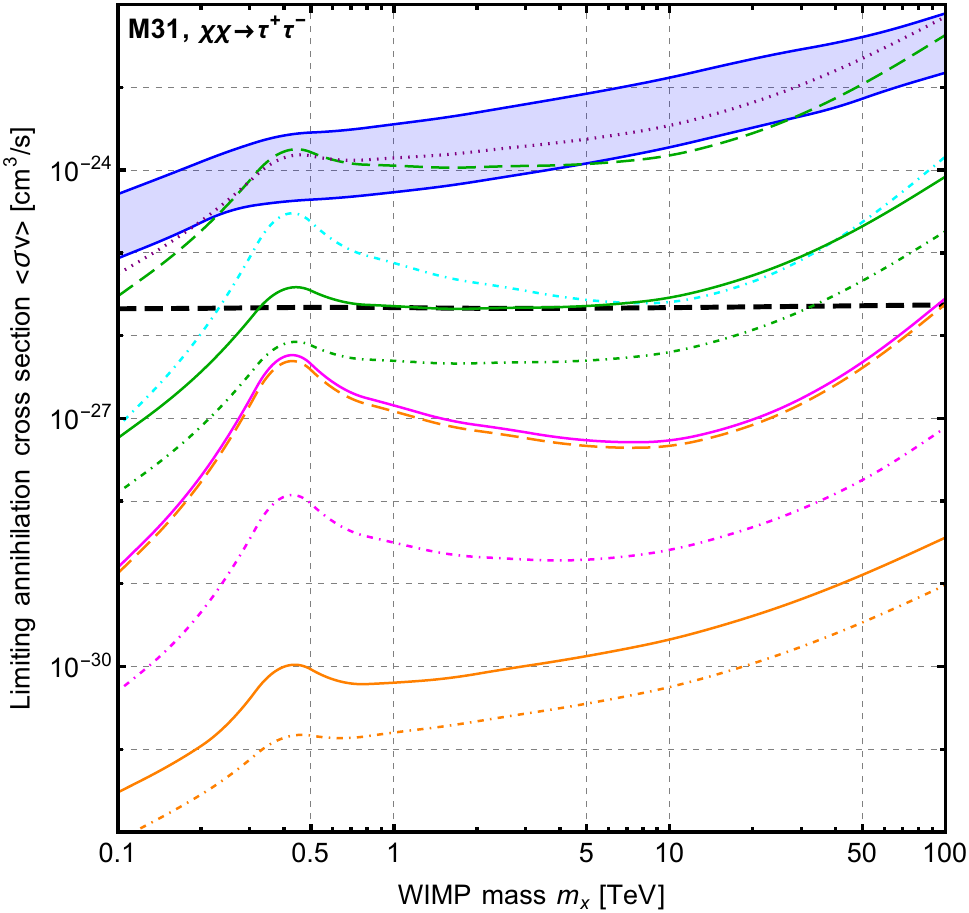}
	\hfill
	\includegraphics[width=0.495\textwidth]{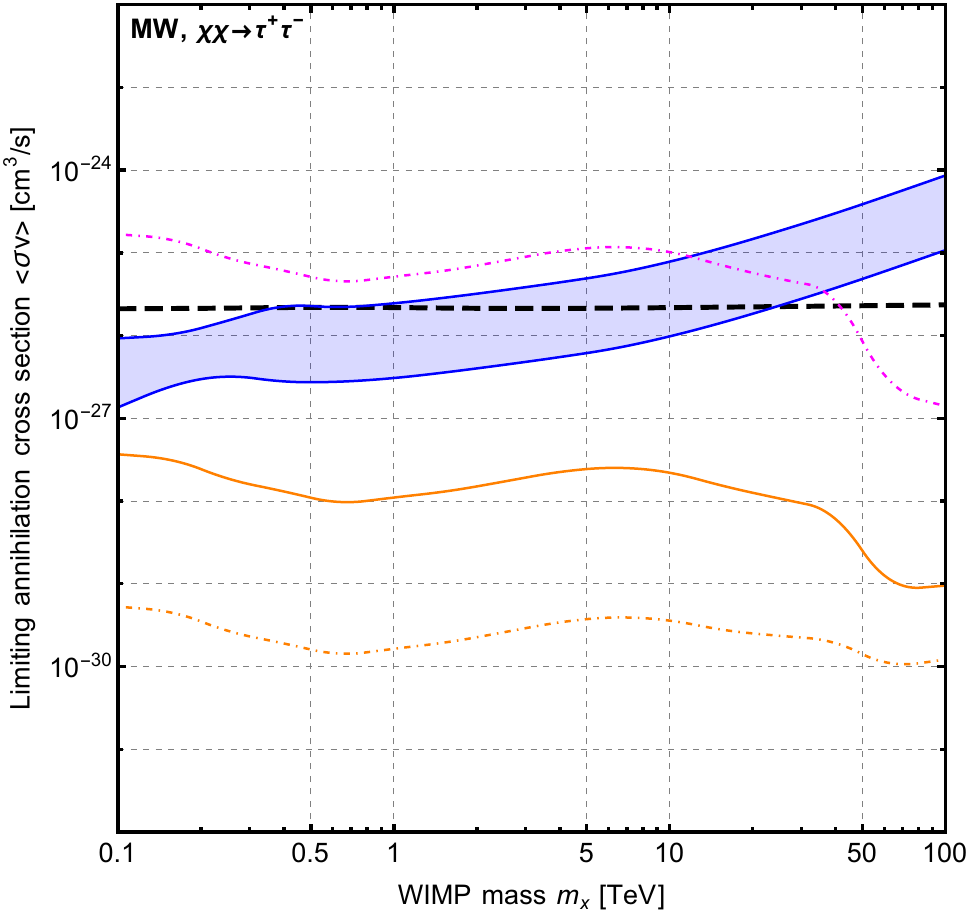}
	\caption{\label{fig:sv}The combined sensitivity limits on annihilating WIMP properties from current Fermi-LAT data and future CTA observations for both considered galaxies and annihilation channels. The blue shaded bands show the constraints from observations of DM galactic halo derived in \cite{2020PhRvD.102d3012A,2021JCAP...01..057A} for MW and in \cite{2019PhRvD..99l3027D,2023JCAP...08..073M} for M31. The bands' width reflects the uncertainties in DM halo density profile. The exclusion lines display the limits from DM density spikes around SMBHs derived here for the listed parameter configurations. The horizontal black dashed line provides the thermal relic annihilation cross section from \cite{2020JCAP...08..011S,sv}.}
\end{figure}

The obtained sensitivity plots are in figure \ref{fig:sv}, they present the main results of this work. In order to assess the sensitivity progress provided by DM spike, at first I collected the current Fermi-LAT and future CTA constraints coming from the observations of diffuse emission from DM galactic halo. These constraints are displayed by the blue stripes, the references to their origin are given in figure caption. The stipes encloses approximately the uncertainty range due to that in DM halo density distributions. This uncertainty range sweeps typically almost an order of magnitude in cross section. Majority of exclusion lines in figure \ref{fig:sv} has quite well visible bump. It corresponds to the transition WIMP mass, where CTA becomes more sensitive than Fermi-LAT. In some cases this bump may be artificially strong just due to coarseness of mass grid. We see that M31 halo provides much weaker constraints than MW halo -- the former does not reach the thermal cross section at any WIMP masses of interest. As was already pointed out in section \ref{sec:i}, MW halo allows to exclude $m_x\lesssim(0.3-10)$ TeV depending on the channel and density profile. Now let us look at the derived sensitivity limits from DM spikes. They are computed for several representative parameter configurations, which add at least some progress to the halo limits. Some numerical values of the spike model parameters are presented in table \ref{tab:2}. In general, 9 independent configurations exist for any $\gamma$ value: $3[b\leftrightarrow R_{sp}] \times 3[\rho_0=\rho(R_{sp})]$. But to be concise, table \ref{tab:2} contains only the "diagonal" values of $\rho_0$ uncertainty "matrix" for each galaxy. Considering MW* limits, we see that only extremely dense and large spike with $\gamma\approx2.3$ and $b\gtrsim0.6$ \emph{together} with quite dense MED--MAX external halo will yield significant progress, i.e. possible exclusion of all thermal WIMPs. But so optimistic spike scenario has a very low probability. And realistic spikes are completely useless in comparison with the smooth halo. Having this, we have to keep in mind, that these constraints are based on the conservative assumption of no further progress in J1745--290 imaging, as was discussed in the beginning of this section. The derived MW* exclusion limits roughly agree with those in \cite{2023JCAP...08..063B}. But the latter work employed different $\rho_{ext}(R)$ profiles. Hence, direct comparison of the results is not meaningful.

M31* demonstrates much richer possibilities, i.e. much larger part of the spike parameter space provides the limits below both the thermal cross section and the halo limits. Thus, a major fraction of adiabatic configurations with $\gamma\approx2.3$ excludes all WIMP masses! Some meaningful exclusion exists even for MIN halo profile in this case (shown by the cyan dot-dashed line). Moreover, a somewhat realistic spike with $\gamma\approx2.0$ yields useful constraints for a significant range of halo densities and spike radii. And only Bahcall-Wolf spike with $\gamma\approx1.75$ merges the limits with those from the halo, which matches our expectations from figure \ref{fig:J}. Fermi-LAT defines the constraints in the mass range up to $\approx$ 5 TeV for $b\bar{b}$ channel and $\approx$ 0.4 TeV for $\tau^+\tau^-$ channel. CTA will take over at higher masses. Such difference between channels is caused by that in the signal spectra $dN_\gamma/dE$: $b\bar{b}$ spectrum peaks at much smaller photon energies than $\tau^+\tau^-$, hence the latter is less available for Fermi-LAT.

Now let us discuss some general features and caveats of the obtained results. The spike slope $\gamma$ has the strongest influence on $J$-factor and the sensitivity. The next model parameter by influence is $\rho_0$. And the spike size parameter $b$ makes probably the smallest impact. Nevertheless, we can notice a very significant sensitivity loss, when $b$ goes from MED value 0.6 down to the lowered value 0.2. One such case is reflected by the orange continuous and dashed lines respectively in figure \ref{fig:sv}. Overall, WIMP searches in SMBH spike suffer from the uncertainties, which are quite similar to those in the case of galactic halo: only steep spike and halo profiles provide good sensitivity. Also we may note, that the dependence of limiting cross section on mass is more peculiar for the spike in comparison with other targets. Such difference arises in the case of spike annihilation core presence and can be understood from eq. \eqref{eq:F}: for usual targets $J(\langle\sigma v\rangle,m_x)=$ const. And another important aspect is that CTA sensitivity depends on exposure time. Figure \ref{fig:sv} reflects the particular choice $T$ = 100 h. In reality the latter could be slightly larger for the same sensitivity due to simplicity of my methodology, as was discussed above. Nevertheless, the point source sensitivity (minimally detectable flux) scales with exposure time as $T^{-1\div-0.5}$. And considering a great importance of M31* for WIMP indirect searches, it could be sensible to dedicate more observational time to this target. Thus, if we would increase the time by $\varDelta$ times, then we would expect the decrease of available cross section values by $\sqrt{\varDelta}\div\varDelta$ times (very approximately). Moreover, the most interesting highest WIMP masses correspond to majorly background-free regime of detection, when CTA sensitivity is limited by the number of signal photons only and has the strongest, i.e. hyperbolic, dependence on exposure. Hence, the heaviest WIMPs will "benefit" the most from an increase of $T$.

Various limits in figure \ref{fig:sv} are not exactly equivalent. Thus, Fermi-LAT and CTA halo limits derived in \cite{2020PhRvD.102d3012A,2021JCAP...01..057A,2019PhRvD..99l3027D,2023JCAP...08..073M} might be based on slightly different halo density profiles (for same galaxy). Also my spike limits have approximately 2$\sigma_s$ significance for MW*, but 5$\sigma_s$ for M31*. However all such minor systematical and statistical uncertainties have a secondary role in comparison with the main systematics of the spike model. Thus, even the presented rough and somewhat qualitative comparison allows to outline all the important implications for WIMP detection strategy.

\section{\label{sec:con}Conclusions and discussion}

This work conducted a comprehensive study of s-wave annihilating WIMP detection prospects in DM density spikes around nearby SMBHs. Conventional spike models were employed from previous works on this subject. An especial emphasis was made on heavy TeV-scale WIMPs in the context of upcoming CTA operation. The main motivation behind my work is to find out a potential way to fully test such WIMPs, since all other developed methodologies reach reliably just $m_x\lesssim0.1$ TeV now and will achieve $m_x\lesssim10$ TeV in future in a good case scenario. Thus, I aimed to identify without a bias the most promising SMBHs and estimate their potential for WIMP detection. The following main conclusions have been deduced.
\begin{enumerate}
	\item SMBH observations at very high gamma-ray energies may probe the heaviest WIMPs with $m_x\sim(10-100)$ TeV. Combination with other probes may allow to search the basic thermal s-wave annihilating WIMPs over the \emph{whole} possible mass range up to the theoretical limit $\approx$ 100 TeV, discovering or excluding such WIMPs as a viable DM candidate.
	\item Many previous studies of DM spike implications likely underestimated the spike radius by $\approx$ 3 times, which lowers the expected emission flux from spike significantly.
	\item Among all nearby relevant SMBHs, only MW* and M31* may manifest some detectable emission flux due to DM for modern gamma-ray instruments. Other objects, including M87*, are not worthwhile for consideration in our context.
	\item The most realistic spike with density profile slope $\gamma=1.5$, which seems to be a consensus among theoretical models, are not detectable at all in any SMBHs, since its brightness is below that of DM galactic halo.
	\item Considering optimistic spike configurations for MW* and M31*, these objects deserve an equal level of attention. MW* was predicted to be slightly brighter, but M31* has much fainter astrophysical background emissions. And overall winner is unpredictable currently. Thus, if CTA will not resolve in details the bright mixed source J1745--290, which contains MW*, then the latter will not be able to add any progress to WIMP sensitivity from the Galactic halo observations, unless the spike possesses extreme and very unlikely density configuration (see figure \ref{fig:sv}).
	\item M31* provides much better sensitivity due to absence of nuisance emissions: thermal WIMPs with any mass will be detectable by CTA for a major fraction of adiabatic spike configurations. Certain WIMP masses will be reachable even for some more realistic spike configurations with $\gamma\approx2.0$. Fermi-LAT data already excludes a substantial part of parameter space at the masses $m_x\lesssim(0.4-5)$ TeV. M31* adds a big sensitivity progress with respect to M31 halo observations. Therefore, I suggest to dedicate larger possible observational time at CTA to this fascinating object (e.g. 200--300 h, comparable to that for MW*).
	\item WIMP sensitivity limits from spike suffer even more from uncertainties in spike density profile, than the limits from halo suffer due to ambiguity in halo profile. And the latter uncertainty influences simultaneously both types of limits, since the halo profile defines the spike base density $\rho_0$. Thus, it appears impossible to evade by anyhow this fundamental systematics.
\end{enumerate}

A detection of any DM spike candidate would need further verification through measurement of its spectrum. The latter requires differential detection in multiple narrow energy bins, which in turn demands significantly larger exposure in comparison with the minimal broadband detection explored here. In this view, the derived sensitivity limits represent the biggest possible potential. Of course, a detection or non-detection of M31 by CTA can not be interpreted as presence or absence of WIMPs. Much deeper investigation will be needed. And other highly advanced gamma-ray telescopes planned for future may facilitate further progress too, indeed: e.g., SWGO \cite{SWGO}, AMS-100 \cite{AMS-100}, HERD \cite{HERD}, GAMMA-400 \cite{2018PAN....81..373E}.

Finally, I neglected in my estimations by relativistic effects near SMBH, particularly by p-wave velocity dependent contribution to the annihilation cross section. For a rough validity check of this approximation, let us estimate the distance from SMBH, where the particle speed reaches 10\% of speed of light: $R(v=0.1c)\approx4\cdot10^{-5}R_{sp}(b=0.6)$ (for M31*). Thus, the volume inside this radius represents a tiny fraction of the whole spike volume. Hence, the non-relativistic treatment should be reasonable.

\acknowledgments
I acknowledge the use of Wolfram Mathematica \circledR~and WebPlotDigitizer \cite{WPD} software, as well as HyperLeda \cite{2014A&A...570A..13M,HyperLeda} and NED \cite{NED} databases.

\bibliography{../../../universal}

\providecommand{\href}[2]{#2}\begingroup\raggedright\begin{thebibliography}{10}

\bibitem{1937ApJ....86..217Z}
F.~{Zwicky}, \emph{{On the Masses of Nebulae and of Clusters of Nebulae}},
  \href{https://doi.org/10.1086/143864}{\emph{Astrophys. J.} {\bfseries 86}
  (1937) 217}.

\bibitem{1936ApJ....83...23S}
S.~{Smith}, \emph{{The Mass of the Virgo Cluster}},
  \href{https://doi.org/10.1086/143697}{\emph{Astrophys. J.} {\bfseries 83}
  (1936) 23}.

\bibitem{2024arXiv240601705C}
M.~{Cirelli}, A.~{Strumia} and J.~{Zupan}, \emph{{Dark Matter}},
  \href{https://doi.org/10.48550/arXiv.2406.01705}{\emph{arXiv} (2024)
  arXiv:2406.01705} [\href{https://arxiv.org/abs/2406.01705}{{\ttfamily
  2406.01705}}].

\bibitem{1984NuPhB.238..453E}
J.~{Ellis}, J.S.~{Hagelin}, D.V.~{Nanopoulos}, K.~{Olive} and M.~{Srednicki},
  \emph{{Supersymmetric relics from the big bang}},
  \href{https://doi.org/10.1016/0550-3213(84)90461-9}{\emph{Nucl. Phys. B}
  {\bfseries 238} (1984) 453}.

\bibitem{2022ScPP...12..163C}
F.~{Calore}, M.~{Cirelli}, L.~{Derome}, Y.~{Genolini}, D.~{Maurin}, P.~{Salati}
  et~al., \emph{{AMS-02 antiprotons and dark matter: Trimmed hints and robust
  bounds}}, \href{https://doi.org/10.21468/SciPostPhys.12.5.163}{\emph{SciPost
  Phys.} {\bfseries 12} (2022) 163}
  [\href{https://arxiv.org/abs/2202.03076}{{\ttfamily 2202.03076}}].

\bibitem{2025arXiv250820229L}
{LAT Collaboration}, S.~{Abdollahi}, L.~{Baldini}, R.~{Bellazzini},
  B.~{Berenji}, E.~{Bissaldi} et~al., \emph{{Combined dark matter search
  towards dwarf spheroidal galaxies with Fermi-LAT, HAWC, H.E.S.S., MAGIC, and
  VERITAS}}, \href{https://doi.org/10.48550/arXiv.2508.20229}{\emph{arXiv}
  (2025) arXiv:2508.20229} [\href{https://arxiv.org/abs/2508.20229}{{\ttfamily
  2508.20229}}].

\bibitem{2023ecrs.confE.120E}
A.~{Egorov}, \emph{{Updated constraints on WIMP dark matter annihilation by
  radio observations of M31 - all annihilation channels}},  in \emph{27th
  European Cosmic Ray Symposium}, p.~120, Jan., 2023,
  \href{https://doi.org/10.22323/1.423.0120}{DOI}
  [\href{https://arxiv.org/abs/2208.14186}{{\ttfamily 2208.14186}}].

\bibitem{2019PhRvD.100d3029S}
J.~{Smirnov} and J.F.~{Beacom}, \emph{{Tev-scale thermal WIMPs: Unitarity and
  its consequences}},
  \href{https://doi.org/10.1103/PhysRevD.100.043029}{\emph{Phys. Rev. D}
  {\bfseries 100} (2019) 043029}
  [\href{https://arxiv.org/abs/1904.11503}{{\ttfamily 1904.11503}}].

\bibitem{2021JCAP...01..057A}
A.~{Acharyya}, R.~{Adam}, C.~{Adams}, I.~{Agudo}, A.~{Aguirre-Santaella},
  R.~{Alfaro} et~al., \emph{{Sensitivity of the Cherenkov Telescope Array to a
  dark matter signal from the Galactic centre}},
  \href{https://doi.org/10.1088/1475-7516/2021/01/057}{\emph{{J. Cosmol.
  Astropart. Phys.}} {\bfseries 01} (2021) 057}
  [\href{https://arxiv.org/abs/2007.16129}{{\ttfamily 2007.16129}}].

\bibitem{2025arXiv250608084A}
S.~{Abe}, T.~{Inada}, E.~{Moulin}, N.L.~{Rodd}, B.R.~{Safdi} and W.L.~{Xu},
  \emph{{Discovering the Higgsino at CTAO-North within the Decade}},
  \href{https://doi.org/10.48550/arXiv.2506.08084}{\emph{arXiv} (2025)
  arXiv:2506.08084} [\href{https://arxiv.org/abs/2506.08084}{{\ttfamily
  2506.08084}}].

\bibitem{2019JCAP...12..061V}
A.~{Viana}, H.~{Schoorlemmer}, A.~{Albert}, V.~{de Souza}, J.P.~{Harding} and
  J.~{Hinton}, \emph{{Searching for dark matter in the Galactic halo with a
  wide field of view TeV gamma-ray observatory in the Southern Hemisphere}},
  \href{https://doi.org/10.1088/1475-7516/2019/12/061}{\emph{{J. Cosmol.
  Astropart. Phys.}} {\bfseries 12} (2019) 061}
  [\href{https://arxiv.org/abs/1906.03353}{{\ttfamily 1906.03353}}].

\bibitem{2019PhRvD..99l3017H}
N.~{Hiroshima}, M.~{Hayashida} and K.~{Kohri}, \emph{{Dependence of accessible
  dark matter annihilation cross sections on the density profiles of dwarf
  spheroidal galaxies with the Cherenkov Telescope Array}},
  \href{https://doi.org/10.1103/PhysRevD.99.123017}{\emph{Phys. Rev. D}
  {\bfseries 99} (2019) 123017}
  [\href{https://arxiv.org/abs/1905.12940}{{\ttfamily 1905.12940}}].

\bibitem{2023JCAP...08..073M}
M.~{Michailidis}, L.~{Marafatto}, D.~{Malyshev}, F.~{Iocco}, G.~{Zaharijas},
  O.~{Sergijenko} et~al., \emph{{Prospects for annihilating dark matter from
  M31 and M33 observations with the Cherenkov Telescope Array}},
  \href{https://doi.org/10.1088/1475-7516/2023/08/073}{\emph{{J. Cosmol.
  Astropart. Phys.}} {\bfseries 08} (2023) 073}
  [\href{https://arxiv.org/abs/2304.08202}{{\ttfamily 2304.08202}}].

\bibitem{1999PhRvL..83.1719G}
P.~{Gondolo} and J.~{Silk}, \emph{{Dark Matter Annihilation at the Galactic
  Center}}, \href{https://doi.org/10.1103/PhysRevLett.83.1719}{\emph{Phys. Rev.
  Lett.} {\bfseries 83} (1999) 1719}
  [\href{https://arxiv.org/abs/astro-ph/9906391}{{\ttfamily
  astro-ph/9906391}}].

\bibitem{2004PhRvL..92t1304M}
D.~{Merritt}, \emph{{Evolution of the Dark Matter Distribution at the Galactic
  Center}}, \href{https://doi.org/10.1103/PhysRevLett.92.201304}{\emph{Phys.
  Rev. Lett.} {\bfseries 92} (2004) 201304}
  [\href{https://arxiv.org/abs/astro-ph/0311594}{{\ttfamily
  astro-ph/0311594}}].

\bibitem{2004PhRvL..93f1302G}
O.Y.~{Gnedin} and J.R.~{Primack}, \emph{{Dark Matter Profile in the Galactic
  Center}}, \href{https://doi.org/10.1103/PhysRevLett.93.061302}{\emph{Phys.
  Rev. Lett.} {\bfseries 93} (2004) 061302}
  [\href{https://arxiv.org/abs/astro-ph/0308385}{{\ttfamily
  astro-ph/0308385}}].

\bibitem{2007PhRvD..76j3532V}
E.~{Vasiliev}, \emph{{Dark matter annihilation near a black hole: Plateau
  versus weak cusp}},
  \href{https://doi.org/10.1103/PhysRevD.76.103532}{\emph{Phys. Rev. D}
  {\bfseries 76} (2007) 103532}
  [\href{https://arxiv.org/abs/0707.3334}{{\ttfamily 0707.3334}}].

\bibitem{2013PhRvD..88f3522S}
L.~{Sadeghian}, F.~{Ferrer} and C.M.~{Will}, \emph{{Dark-matter distributions
  around massive black holes: A general relativistic analysis}},
  \href{https://doi.org/10.1103/PhysRevD.88.063522}{\emph{Phys. Rev. D}
  {\bfseries 88} (2013) 063522}
  [\href{https://arxiv.org/abs/1305.2619}{{\ttfamily 1305.2619}}].

\bibitem{2016PhRvD..93l3510S}
S.L.~{Shapiro} and J.~{Shelton}, \emph{{Weak annihilation cusp inside the dark
  matter spike about a black hole}},
  \href{https://doi.org/10.1103/PhysRevD.93.123510}{\emph{Phys. Rev. D}
  {\bfseries 93} (2016) 123510}
  [\href{https://arxiv.org/abs/1606.01248}{{\ttfamily 1606.01248}}].

\bibitem{2022PhRvD.106d3018S}
S.L.~{Shapiro} and D.C.~{Heggie}, \emph{{Effect of stars on the dark matter
  spike around a black hole: A tale of two treatments}},
  \href{https://doi.org/10.1103/PhysRevD.106.043018}{\emph{Phys. Rev. D}
  {\bfseries 106} (2022) 043018}
  [\href{https://arxiv.org/abs/2209.08105}{{\ttfamily 2209.08105}}].

\bibitem{2025PhRvD.112d3025Z}
Z.-C.~{Zhang}, H.-C.~{Yuan} and Y.~{Tang}, \emph{{Universal density and
  velocity distributions of dark matter around massive black holes}},
  \href{https://doi.org/10.1103/21h1-kmwq}{\emph{Phys. Rev. D} {\bfseries 112}
  (2025) 043025} [\href{https://arxiv.org/abs/2503.02573}{{\ttfamily
  2503.02573}}].

\bibitem{2024MNRAS.527.3196S}
Z.-Q.~{Shen}, G.-W.~{Yuan}, C.-Z.~{Jiang}, Y.-L.S.~{Tsai}, Q.~{Yuan} and
  Y.-Z.~{Fan}, \emph{{Exploring dark matter spike distribution around the
  Galactic centre with stellar orbits}},
  \href{https://doi.org/10.1093/mnras/stad3282}{\emph{Mon. Not. R. Astron.
  Soc.} {\bfseries 527} (2024) 3196}
  [\href{https://arxiv.org/abs/2303.09284}{{\ttfamily 2303.09284}}].

\bibitem{2025arXiv250610122S}
M.~{Sharma}, G.~{Herrera}, N.~{Arav} and S.~{Horiuchi}, \emph{{A novel method
  to trace the dark matter density profile around supermassive black holes with
  AGN reverberation mapping}},
  \href{https://doi.org/10.48550/arXiv.2506.10122}{\emph{arXiv} (2025)
  arXiv:2506.10122} [\href{https://arxiv.org/abs/2506.10122}{{\ttfamily
  2506.10122}}].

\bibitem{2014PhRvL.113o1302F}
B.D.~{Fields}, S.L.~{Shapiro} and J.~{Shelton}, \emph{{Galactic Center
  Gamma-Ray Excess from Dark Matter Annihilation: Is There a Black Hole
  Spike?}}, \href{https://doi.org/10.1103/PhysRevLett.113.151302}{\emph{Phys.
  Rev. Lett.} {\bfseries 113} (2014) 151302}
  [\href{https://arxiv.org/abs/1406.4856}{{\ttfamily 1406.4856}}].

\bibitem{2015PhRvD..92d3510L}
T.~{Lacroix}, C.~{B{\r{A}}`hm} and J.~{Silk}, \emph{{Ruling out thermal dark
  matter with a black hole induced spiky profile in the M87 galaxy}},
  \href{https://doi.org/10.1103/PhysRevD.92.043510}{\emph{Phys. Rev. D}
  {\bfseries 92} (2015) 043510}
  [\href{https://arxiv.org/abs/1505.00785}{{\ttfamily 1505.00785}}].

\bibitem{2015PhRvL.115w1302S}
J.~{Shelton}, S.L.~{Shapiro} and B.D.~{Fields}, \emph{{Black Hole Window into
  p-Wave Dark Matter Annihilation}},
  \href{https://doi.org/10.1103/PhysRevLett.115.231302}{\emph{Phys. Rev. Lett.}
  {\bfseries 115} (2015) 231302}
  [\href{https://arxiv.org/abs/1506.04143}{{\ttfamily 1506.04143}}].

\bibitem{2020PhRvD.102b3030C}
B.T.~{Chiang}, S.L.~{Shapiro} and J.~{Shelton}, \emph{{Faint dark matter
  annihilation signals and the Milky Way's supermassive black hole}},
  \href{https://doi.org/10.1103/PhysRevD.102.023030}{\emph{Phys. Rev. D}
  {\bfseries 102} (2020) 023030}
  [\href{https://arxiv.org/abs/1912.09446}{{\ttfamily 1912.09446}}].

\bibitem{2023PhRvD.108j3042C}
K.~{Christy}, J.~{Kumar} and P.~{Sandick}, \emph{{Constraining p -wave dark
  matter annihilation with gamma-ray observations of M87}},
  \href{https://doi.org/10.1103/PhysRevD.108.103042}{\emph{Phys. Rev. D}
  {\bfseries 108} (2023) 103042}
  [\href{https://arxiv.org/abs/2305.05155}{{\ttfamily 2305.05155}}].

\bibitem{2023JCAP...08..063B}
S.~{Balaji}, D.~{Sachdeva}, F.~{Sala} and J.~{Silk}, \emph{{Dark matter spikes
  around Sgr A* in {\ensuremath{\gamma}}-rays}},
  \href{https://doi.org/10.1088/1475-7516/2023/08/063}{\emph{{J. Cosmol.
  Astropart. Phys.}} {\bfseries 08} (2023) 063}
  [\href{https://arxiv.org/abs/2303.12107}{{\ttfamily 2303.12107}}].

\bibitem{2025PhRvD.111k5033P}
M.~{Phoroutan-Mehr} and H.-B.~{Yu}, \emph{{Relaxing constraints on dark matter
  annihilation near the supermassive black hole in M87}},
  \href{https://doi.org/10.1103/g5cn-64kw}{\emph{Phys. Rev. D} {\bfseries 111}
  (2025) 115033} [\href{https://arxiv.org/abs/2411.18751}{{\ttfamily
  2411.18751}}].

\bibitem{CTA}
\url{https://www.ctao.org/for-scientists/performance/}.

\bibitem{2011ApJ...728L..10L}
Z.~{Li}, M.R.~{Garcia}, W.R.~{Forman}, C.~{Jones}, R.P.~{Kraft}, D.V.~{Lal}
  et~al., \emph{{The Murmur of the Hidden Monster: Chandra's Decadal View of
  the Supermassive Black Hole in M31}},
  \href{https://doi.org/10.1088/2041-8205/728/1/L10}{\emph{Astrophys. J. Lett.}
  {\bfseries 728} (2011) L10}
  [\href{https://arxiv.org/abs/1011.1224}{{\ttfamily 1011.1224}}].

\bibitem{2013PhRvD..88a4035K}
J.~{Kumar} and D.~{Marfatia}, \emph{{Matrix element analyses of dark matter
  scattering and annihilation}},
  \href{https://doi.org/10.1103/PhysRevD.88.014035}{\emph{Phys. Rev. D}
  {\bfseries 88} (2013) 014035}
  [\href{https://arxiv.org/abs/1305.1611}{{\ttfamily 1305.1611}}].

\bibitem{2017PhRvD..96h3014F}
F.~{Ferrer}, A.~{Medeiros da Rosa} and C.M.~{Will}, \emph{{Dark matter spikes
  in the vicinity of Kerr black holes}},
  \href{https://doi.org/10.1103/PhysRevD.96.083014}{\emph{Phys. Rev. D}
  {\bfseries 96} (2017) 083014}
  [\href{https://arxiv.org/abs/1707.06302}{{\ttfamily 1707.06302}}].

\bibitem{2015ApJ...806..264S}
J.D.~{Schnittman}, \emph{{The Distribution and Annihilation of Dark Matter
  Around Black Holes}},
  \href{https://doi.org/10.1088/0004-637X/806/2/264}{\emph{Astrophys. J.}
  {\bfseries 806} (2015) 264}
  [\href{https://arxiv.org/abs/1506.06728}{{\ttfamily 1506.06728}}].

\bibitem{PPPC}
\url{http://www.marcocirelli.net/PPPC4DMID.html}.

\bibitem{2011JCAP...03..051C}
M.~{Cirelli}, G.~{Corcella}, A.~{Hektor}, G.~{H{\"u}tsi}, M.~{Kadastik},
  P.~{Panci} et~al., \emph{{PPPC 4 DM ID: a poor particle physicist cookbook
  for dark matter indirect detection}},
  \href{https://doi.org/10.1088/1475-7516/2011/03/051}{\emph{{J. Cosmol.
  Astropart. Phys.}} {\bfseries 03} (2011) 051}
  [\href{https://arxiv.org/abs/1012.4515}{{\ttfamily 1012.4515}}].

\bibitem{2011JCAP...03..019C}
P.~{Ciafaloni}, D.~{Comelli}, A.~{Riotto}, F.~{Sala}, A.~{Strumia} and
  A.~{Urbano}, \emph{{Weak corrections are relevant for dark matter indirect
  detection}}, \href{https://doi.org/10.1088/1475-7516/2011/03/019}{\emph{{J.
  Cosmol. Astropart. Phys.}} {\bfseries 03} (2011) 019}
  [\href{https://arxiv.org/abs/1009.0224}{{\ttfamily 1009.0224}}].

\bibitem{2010PhRvD..82h3514G}
M.~{Gorchtein}, S.~{Profumo} and L.~{Ubaldi}, \emph{{Probing dark matter with
  active galactic nuclei jets}},
  \href{https://doi.org/10.1103/PhysRevD.82.083514}{\emph{Phys. Rev. D}
  {\bfseries 82} (2010) 083514}
  [\href{https://arxiv.org/abs/1008.2230}{{\ttfamily 1008.2230}}].

\bibitem{2001PhRvD..64d3504U}
P.~{Ullio}, H.~{Zhao} and M.~{Kamionkowski}, \emph{{Dark-matter spike at the
  galactic center?}},
  \href{https://doi.org/10.1103/PhysRevD.64.043504}{\emph{Phys. Rev. D}
  {\bfseries 64} (2001) 043504}
  [\href{https://arxiv.org/abs/astro-ph/0101481}{{\ttfamily
  astro-ph/0101481}}].

\bibitem{2002PhRvL..88s1301M}
D.~{Merritt}, M.~{Milosavljevi{\'c}}, L.~{Verde} and R.~{Jimenez}, \emph{{Dark
  Matter Spikes and Annihilation Radiation from the Galactic Center}},
  \href{https://doi.org/10.1103/PhysRevLett.88.191301}{\emph{Phys. Rev. Lett.}
  {\bfseries 88} (2002) 191301}
  [\href{https://arxiv.org/abs/astro-ph/0201376}{{\ttfamily
  astro-ph/0201376}}].

\bibitem{2016MNRAS.456.3542T}
E.~{Tollet}, A.V.~{Macci{\`o}}, A.A.~{Dutton}, G.S.~{Stinson}, L.~{Wang},
  C.~{Penzo} et~al., \emph{{NIHAO - IV: core creation and destruction in dark
  matter density profiles across cosmic time}},
  \href{https://doi.org/10.1093/mnras/stv2856}{\emph{Mon. Not. R. Astron. Soc.}
  {\bfseries 456} (2016) 3542}
  [\href{https://arxiv.org/abs/1507.03590}{{\ttfamily 1507.03590}}].

\bibitem{2012A&A...546A...4T}
A.~{Tamm}, E.~{Tempel}, P.~{Tenjes}, O.~{Tihhonova} and T.~{Tuvikene},
  \emph{{Stellar mass map and dark matter distribution in M 31}},
  \href{https://doi.org/10.1051/0004-6361/201220065}{\emph{Astron. Astrophys.}
  {\bfseries 546} (2012) A4} [\href{https://arxiv.org/abs/1208.5712}{{\ttfamily
  1208.5712}}].

\bibitem{2004cbhg.symp..263M}
D.~{Merritt}, \emph{{Single and Binary Black Holes and their Influence on
  Nuclear Structure}},  in \emph{Coevolution of Black Holes and Galaxies},
  L.C.~{Ho}, ed., p.~263, Jan., 2004,
  \href{https://doi.org/10.48550/arXiv.astro-ph/0301257}{DOI}
  [\href{https://arxiv.org/abs/astro-ph/0301257}{{\ttfamily
  astro-ph/0301257}}].

\bibitem{2018A&A...609A..27S}
R.~{Sch{\"o}del}, E.~{Gallego-Cano}, H.~{Dong}, F.~{Nogueras-Lara},
  A.T.~{Gallego-Calvente}, P.~{Amaro-Seoane} et~al., \emph{{The distribution of
  stars around the Milky Way's central black hole. II. Diffuse light from
  sub-giants and dwarfs}},
  \href{https://doi.org/10.1051/0004-6361/201730452}{\emph{Astron. Astrophys.}
  {\bfseries 609} (2018) A27}
  [\href{https://arxiv.org/abs/1701.03817}{{\ttfamily 1701.03817}}].

\bibitem{SMBH}
\url{https://en.wikipedia.org/wiki/List_of_black_holes}.

\bibitem{2022PhRvD.106b3023E}
A.E.~{Egorov}, \emph{{Updated constraints on WIMP dark matter annihilation by
  radio observations of M31}},
  \href{https://doi.org/10.1103/PhysRevD.106.023023}{\emph{Phys. Rev. D}
  {\bfseries 106} (2022) 023023}
  [\href{https://arxiv.org/abs/2205.01033}{{\ttfamily 2205.01033}}].

\bibitem{2013PhRvD..88b3504E}
A.E.~{Egorov} and E.~{Pierpaoli}, \emph{{Constraints on dark matter
  annihilation by radio observations of M31}},
  \href{https://doi.org/10.1103/PhysRevD.88.023504}{\emph{Phys. Rev. D}
  {\bfseries 88} (2013) 023504}
  [\href{https://arxiv.org/abs/1304.0517}{{\ttfamily 1304.0517}}].

\bibitem{Burkert}
A.~{Burkert}, \emph{{The Structure of Dark Matter Halos in Dwarf Galaxies}},
  \href{https://doi.org/10.1086/309560}{\emph{Astrophys. J. Lett.} {\bfseries
  447} (1995) L25} [\href{https://arxiv.org/abs/astro-ph/9504041}{{\ttfamily
  astro-ph/9504041}}].

\bibitem{2014A&A...570A.132S}
S.~{Samurovi{\'c}}, \emph{{Investigation of dark matter and modified Newtonian
  dynamics in early-type galaxies through globular cluster systems}},
  \href{https://doi.org/10.1051/0004-6361/201321459}{\emph{Astron. Astrophys.}
  {\bfseries 570} (2014) A132}.

\bibitem{2025arXiv251002439A}
R.M.~{Al-Amri}, J.L.~{Walsh}, E.R.~{Liepold}, C.-P.~{Ma} and J.E.~{Greene},
  \emph{{Mapping the Stellar Kinematics in the Central 240 Parsecs of M87 with
  the James Webb Space Telescope}},
  \href{https://doi.org/10.48550/arXiv.2510.02439}{\emph{arXiv} (2025)
  arXiv:2510.02439} [\href{https://arxiv.org/abs/2510.02439}{{\ttfamily
  2510.02439}}].

\bibitem{NED}
NASA/IPAC Extragalactic Database \url{https://ned.ipac.caltech.edu/}.

\bibitem{HyperLeda}
HyperLeda database for physics of galaxies
  \url{http://atlas.obs-hp.fr/hyperleda/}.

\bibitem{2022A&A...657L..12G}
{GRAVITY Collaboration}, R.~{Abuter}, N.~{Aimar}, A.~{Amorim}, J.~{Ball},
  M.~{Baub{\"o}ck} et~al., \emph{{Mass distribution in the Galactic Center
  based on interferometric astrometry of multiple stellar orbits}},
  \href{https://doi.org/10.1051/0004-6361/202142465}{\emph{Astron. Astrophys.}
  {\bfseries 657} (2022) L12}
  [\href{https://arxiv.org/abs/2112.07478}{{\ttfamily 2112.07478}}].

\bibitem{2018A&A...616A..83V}
E.~{Valenti}, M.~{Zoccali}, A.~{Mucciarelli}, O.A.~{Gonzalez}, F.~{Surot},
  D.~{Minniti} et~al., \emph{{The central velocity dispersion of the Milky Way
  bulge}}, \href{https://doi.org/10.1051/0004-6361/201832905}{\emph{Astron.
  Astrophys.} {\bfseries 616} (2018) A83}
  [\href{https://arxiv.org/abs/1805.00275}{{\ttfamily 1805.00275}}].

\bibitem{2021ApJ...920...84L}
S.~{Li}, A.G.~{Riess}, M.P.~{Busch}, S.~{Casertano}, L.M.~{Macri} and
  W.~{Yuan}, \emph{{A Sub-2\% Distance to M31 from Photometrically Homogeneous
  Near-infrared Cepheid Period-Luminosity Relations Measured with the Hubble
  Space Telescope}},
  \href{https://doi.org/10.3847/1538-4357/ac1597}{\emph{Astrophys. J.}
  {\bfseries 920} (2021) 84}
  [\href{https://arxiv.org/abs/2107.08029}{{\ttfamily 2107.08029}}].

\bibitem{2005ApJ...631..280B}
R.~{Bender}, J.~{Kormendy}, G.~{Bower}, R.~{Green}, J.~{Thomas}, A.C.~{Danks}
  et~al., \emph{{HST STIS Spectroscopy of the Triple Nucleus of M31: Two Nested
  Disks in Keplerian Rotation around a Supermassive Black Hole}},
  \href{https://doi.org/10.1086/432434}{\emph{Astrophys. J.} {\bfseries 631}
  (2005) 280} [\href{https://arxiv.org/abs/astro-ph/0509839}{{\ttfamily
  astro-ph/0509839}}].

\bibitem{2009ApJ...698..198G}
K.~{G{\"u}ltekin}, D.O.~{Richstone}, K.~{Gebhardt}, T.R.~{Lauer},
  S.~{Tremaine}, M.C.~{Aller} et~al., \emph{{The M-{\ensuremath{\sigma}} and
  M-L Relations in Galactic Bulges, and Determinations of Their Intrinsic
  Scatter}},
  \href{https://doi.org/10.1088/0004-637X/698/1/198}{\emph{Astrophys. J.}
  {\bfseries 698} (2009) 198}
  [\href{https://arxiv.org/abs/0903.4897}{{\ttfamily 0903.4897}}].

\bibitem{2017MNRAS.468.3949A}
A.B.~{Alabi}, D.A.~{Forbes}, A.J.~{Romanowsky}, J.P.~{Brodie}, J.~{Strader},
  J.~{Janz} et~al., \emph{{The SLUGGS survey: dark matter fractions at large
  radii and assembly epochs of early-type galaxies from globular cluster
  kinematics}}, \href{https://doi.org/10.1093/mnras/stx678}{\emph{Mon. Not. R.
  Astron. Soc.} {\bfseries 468} (2017) 3949}
  [\href{https://arxiv.org/abs/1701.05904}{{\ttfamily 1701.05904}}].

\bibitem{2022AJ....163..234K}
I.D.~{Karachentsev}, L.N.~{Makarova}, G.S.~{Anand} and R.B.~{Tully},
  \emph{{Around the Spindle Galaxy: The Dark Halo Mass of NGC 3115}},
  \href{https://doi.org/10.3847/1538-3881/ac5ab5}{\emph{Astron. J.} {\bfseries
  163} (2022) 234} [\href{https://arxiv.org/abs/2203.01700}{{\ttfamily
  2203.01700}}].

\bibitem{2024ApJ...965..128Y}
X.~{Yan}, R.-S.~{Lu}, W.~{Jiang}, T.P.~{Krichbaum}, F.-G.~{Xie} and
  Z.-Q.~{Shen}, \emph{{Multifrequency Very Long Baseline Interferometry Imaging
  of the Subparsec-scale Jet in the Sombrero Galaxy (M104)}},
  \href{https://doi.org/10.3847/1538-4357/ad31a2}{\emph{Astrophys. J.}
  {\bfseries 965} (2024) 128}
  [\href{https://arxiv.org/abs/2403.04215}{{\ttfamily 2403.04215}}].

\bibitem{2011ApJ...739...21J}
J.R.~{Jardel}, K.~{Gebhardt}, J.~{Shen}, D.B.~{Fisher}, J.~{Kormendy},
  J.~{Kinzler} et~al., \emph{{Orbit-based Dynamical Models of the Sombrero
  Galaxy (NGC 4594)}},
  \href{https://doi.org/10.1088/0004-637X/739/1/21}{\emph{Astrophys. J.}
  {\bfseries 739} (2011) 21} [\href{https://arxiv.org/abs/1107.1238}{{\ttfamily
  1107.1238}}].

\bibitem{2019ApJ...875L...6E}
{Event Horizon Telescope Collaboration}, K.~{Akiyama}, A.~{Alberdi}, W.~{Alef},
  K.~{Asada}, R.~{Azulay} et~al., \emph{{First M87 Event Horizon Telescope
  Results. VI. The Shadow and Mass of the Central Black Hole}},
  \href{https://doi.org/10.3847/2041-8213/ab1141}{\emph{Astrophys. J. Lett.}
  {\bfseries 875} (2019) L6}
  [\href{https://arxiv.org/abs/1906.11243}{{\ttfamily 1906.11243}}].

\bibitem{2011ApJ...729..129M}
J.D.~{Murphy}, K.~{Gebhardt} and J.J.~{Adams}, \emph{{Galaxy Kinematics with
  VIRUS-P: The Dark Matter Halo of M87}},
  \href{https://doi.org/10.1088/0004-637X/729/2/129}{\emph{Astrophys. J.}
  {\bfseries 729} (2011) 129}
  [\href{https://arxiv.org/abs/1101.1957}{{\ttfamily 1101.1957}}].

\bibitem{2022ApJ...929...17D}
M.~{De Laurentis} and P.~{Salucci}, \emph{{The Accurate Mass Distribution of
  M87, the Giant Galaxy with Imaged Shadow of Its Supermassive Black Hole, as a
  Portal to New Physics}},
  \href{https://doi.org/10.3847/1538-4357/ac54b9}{\emph{Astrophys. J.}
  {\bfseries 929} (2022) 17}.

\bibitem{2012ApJ...756..179M}
N.J.~{McConnell}, C.-P.~{Ma}, J.D.~{Murphy}, K.~{Gebhardt}, T.R.~{Lauer},
  J.R.~{Graham} et~al., \emph{{Dynamical Measurements of Black Hole Masses in
  Four Brightest Cluster Galaxies at 100 Mpc}},
  \href{https://doi.org/10.1088/0004-637X/756/2/179}{\emph{Astrophys. J.}
  {\bfseries 756} (2012) 179}
  [\href{https://arxiv.org/abs/1203.1620}{{\ttfamily 1203.1620}}].

\bibitem{2007MNRAS.382..657T}
J.~{Thomas}, R.P.~{Saglia}, R.~{Bender}, D.~{Thomas}, K.~{Gebhardt},
  J.~{Magorrian} et~al., \emph{{Dynamical modelling of luminous and dark matter
  in 17 Coma early-type galaxies}},
  \href{https://doi.org/10.1111/j.1365-2966.2007.12434.x}{\emph{Mon. Not. R.
  Astron. Soc.} {\bfseries 382} (2007) 657}
  [\href{https://arxiv.org/abs/0709.0691}{{\ttfamily 0709.0691}}].

\bibitem{2016Natur.531..476H}
{HESS Collaboration}, A.~{Abramowski}, F.~{Aharonian}, F.A.~{Benkhali},
  A.G.~{Akhperjanian}, E.O.~{Ang{\"u}ner} et~al., \emph{{Acceleration of
  petaelectronvolt protons in the Galactic Centre}},
  \href{https://doi.org/10.1038/nature17147}{\emph{Nature (London)} {\bfseries
  531} (2016) 476} [\href{https://arxiv.org/abs/1603.07730}{{\ttfamily
  1603.07730}}].

\bibitem{2021ApJ...913..115A}
C.B.~{Adams}, W.~{Benbow}, A.~{Brill}, R.~{Brose}, M.~{Buchovecky},
  M.~{Capasso} et~al., \emph{{VERITAS Observations of the Galactic Center
  Region at Multi-TeV Gamma-Ray Energies}},
  \href{https://doi.org/10.3847/1538-4357/abf926}{\emph{Astrophys. J.}
  {\bfseries 913} (2021) 115}
  [\href{https://arxiv.org/abs/2104.12735}{{\ttfamily 2104.12735}}].

\bibitem{2020ApJ...893...16A}
A.~{Albert}, R.~{Alfaro}, C.~{Alvarez}, J.C.~{Arteaga-Vel{\'a}zquez},
  K.P.~{Arunbabu}, D.~{Avila Rojas} et~al., \emph{{Constraints on the Emission
  of Gamma-Rays from M31 with HAWC}},
  \href{https://doi.org/10.3847/1538-4357/ab7999}{\emph{Astrophys. J.}
  {\bfseries 893} (2020) 16}
  [\href{https://arxiv.org/abs/2001.04065}{{\ttfamily 2001.04065}}].

\bibitem{2023ApJ...945L..22X}
Y.~{Xing}, Z.~{Wang}, D.~{Zheng} and J.~{Li}, \emph{{On the Gamma-Ray Emission
  of the Andromeda Galaxy M31}},
  \href{https://doi.org/10.3847/2041-8213/acbf4f}{\emph{Astrophys. J. Lett.}
  {\bfseries 945} (2023) L22}
  [\href{https://arxiv.org/abs/2301.06743}{{\ttfamily 2301.06743}}].

\bibitem{1983ApJ...272..317L}
T.-P.~{Li} and Y.-Q.~{Ma}, \emph{{Analysis methods for results in gamma-ray
  astronomy.}}, \href{https://doi.org/10.1086/161295}{\emph{Astrophys. J.}
  {\bfseries 272} (1983) 317}.

\bibitem{2013APh....43..171B}
K.~{Bernl{\"o}hr}, A.~{Barnacka}, Y.~{Becherini}, O.~{Blanch Bigas},
  E.~{Carmona}, P.~{Colin} et~al., \emph{{Monte Carlo design studies for the
  Cherenkov Telescope Array}},
  \href{https://doi.org/10.1016/j.astropartphys.2012.10.002}{\emph{Astropart.
  Phys.} {\bfseries 43} (2013) 171}
  [\href{https://arxiv.org/abs/1210.3503}{{\ttfamily 1210.3503}}].

\bibitem{2006ApJ...648..890M}
D.~{Merritt} and A.~{Szell}, \emph{{Dynamical Cusp Regeneration}},
  \href{https://doi.org/10.1086/506010}{\emph{Astrophys. J.} {\bfseries 648}
  (2006) 890} [\href{https://arxiv.org/abs/astro-ph/0510498}{{\ttfamily
  astro-ph/0510498}}].

\bibitem{2020PhRvD.102d3012A}
K.N.~{Abazajian}, S.~{Horiuchi}, M.~{Kaplinghat}, R.E.~{Keeley} and
  O.~{Macias}, \emph{{Strong constraints on thermal relic dark matter from
  Fermi-LAT observations of the Galactic Center}},
  \href{https://doi.org/10.1103/PhysRevD.102.043012}{\emph{Phys. Rev. D}
  {\bfseries 102} (2020) 043012}
  [\href{https://arxiv.org/abs/2003.10416}{{\ttfamily 2003.10416}}].

\bibitem{2019PhRvD..99l3027D}
M.~{Di Mauro}, X.~{Hou}, C.~{Eckner}, G.~{Zaharijas} and E.~{Charles},
  \emph{{Search for {\ensuremath{\gamma}}-ray emission from dark matter
  particle interactions from the Andromeda and Triangulum galaxies with the
  Fermi Large Area Telescope}},
  \href{https://doi.org/10.1103/PhysRevD.99.123027}{\emph{Phys. Rev. D}
  {\bfseries 99} (2019) 123027}
  [\href{https://arxiv.org/abs/1904.10977}{{\ttfamily 1904.10977}}].

\bibitem{2020JCAP...08..011S}
K.~{Saikawa} and S.~{Shirai}, \emph{{Precise WIMP dark matter abundance and
  Standard Model thermodynamics}},
  \href{https://doi.org/10.1088/1475-7516/2020/08/011}{\emph{{J. Cosmol.
  Astropart. Phys.}} {\bfseries 08} (2020) 011}
  [\href{https://arxiv.org/abs/2005.03544}{{\ttfamily 2005.03544}}].

\bibitem{sv}
\url{https://member.ipmu.jp/satoshi.shirai/DM2020/}.

\bibitem{SWGO}
{SWGO Collaboration}, P.~{Abreu}, R.~{Alfaro}, A.~{Alfonso}, M.~{Andrade},
  E.O.~{Ang{\"u}ner} et~al., \emph{{Science Prospects for the Southern
  Wide-field Gamma-ray Observatory: SWGO}},
  \href{https://doi.org/10.48550/arXiv.2506.01786}{\emph{arXiv} (2025)
  arXiv:2506.01786} [\href{https://arxiv.org/abs/2506.01786}{{\ttfamily
  2506.01786}}].

\bibitem{AMS-100}
S.~{Schael}, A.~{Atanasyan}, J.~{Berdugo}, T.~{Bretz}, M.~{Czupalla},
  B.~{Dachwald} et~al., \emph{{AMS-100: The next generation magnetic
  spectrometer in space - An international science platform for physics and
  astrophysics at Lagrange point 2}},
  \href{https://doi.org/10.1016/j.nima.2019.162561}{\emph{Nucl. Inst. Meth.
  Phys. Res. A} {\bfseries 944} (2019) 162561}
  [\href{https://arxiv.org/abs/1907.04168}{{\ttfamily 1907.04168}}].

\bibitem{HERD}
I.~{Cagnoli}, D.~{Kyratzis} and D.~{Serini}, \emph{{HERD space mission: Probing
  the Galactic Cosmic Ray frontier}},
  \href{https://doi.org/10.1016/j.nima.2024.169788}{\emph{Nucl. Inst. Meth.
  Phys. Res. A} {\bfseries 1068} (2024) 169788}.

\bibitem{2018PAN....81..373E}
A.E.~{Egorov}, A.M.~{Galper}, N.P.~{Topchiev}, A.A.~{Leonov}, S.I.~{Suchkov},
  M.D.~{Kheymits} et~al., \emph{{Detactability of Dark Matter Subhalos by Means
  of the GAMMA-400 Telescope}},
  \href{https://doi.org/10.1134/S1063778818030110}{\emph{Phys. Atom. Nucl.}
  {\bfseries 81} (2018) 373}
  [\href{https://arxiv.org/abs/1710.02492}{{\ttfamily 1710.02492}}].

\bibitem{WPD}
A. Rohatgi, WebPlotDigitizer \url{https://automeris.io}.

\bibitem{2014A&A...570A..13M}
D.~{Makarov}, P.~{Prugniel}, N.~{Terekhova}, H.~{Courtois} and I.~{Vauglin},
  \emph{{HyperLEDA. III. The catalogue of extragalactic distances}},
  \href{https://doi.org/10.1051/0004-6361/201423496}{\emph{Astron. Astrophys.}
  {\bfseries 570} (2014) A13}.

\end{thebibliography}\endgroup

\end{document}